%% file: template.tex
\documentclass[aps,prb,twocolumn,superscriptaddress]{revtex4-1}
\usepackage[english]{babel}
\usepackage[inline]{enumitem}
\usepackage[T1]{fontenc}
\usepackage{graphicx}
\usepackage{amsmath}
\usepackage{amssymb}
\usepackage{hyperref}
\usepackage{siunitx}
\usepackage{etoolbox}
\usepackage{pdfcomment}
\usepackage{pgfplots}
\usepackage{textcomp}
\usepackage{cleveref}

\crefname{paragraph}{paragraph}{paragraphs}
\Crefname{paragraph}{Paragraph}{Paragraphs}

\usepackage{glossaries}
\newacronym{api}{API}{application programming interface}
\newacronym{dag}{DAG}{Directed Acyclic Graph}
\newacronym{dcg}{DCG}{Directed Cyclic Graph}
\newacronym{gui}{GUI}{graphical user interface}
\newacronym{cwl}{CWL}{Common Workflow Language}
\newacronym{hpc}{HPC}{high-performance computing}
\newacronym{rpc}{RPC}{Remote Procedure Call}
\newacronym[plural=WMS's,firstplural=workflow management systems (WMS's)]{wms}{WMS}{workflow management system}

\definecolor{DARKMAGENTA}{HTML}{af2f40}
\hypersetup{
    colorlinks=true,
    linkcolor=DARKMAGENTA,
    citecolor=DARKMAGENTA,
    urlcolor=DARKMAGENTA,
}

\usepackage{tikz}
\usepackage{tkz-graph}
\usepackage{tikz-uml}
\usetikzlibrary{arrows,automata,calc,positioning,shadows.blur,decorations.pathreplacing,decorations.markings,pgfplots.groupplots}

\tikzstyle{class}=[
    rectangle,
    draw=black,
    text centered,
    anchor=north,
    text=black,
    text width=3cm,
    bottom color=white,
    top color=white]

\usepackage{listings}
\AtBeginDocument{\usepackage{booktabs}}

\begin{document}

\title{Workflows in AiiDA: Engineering a high-throughput, event-based engine for robust and modular computational workflows}
\author{Martin\,Uhrin}
\affiliation{Theory and Simulation of Materials (THEOS), and National Centre for Computational Design and Discovery of Novel Materials (MARVEL), \'{E}cole Polytechnique F\'{e}d\'{e}rale de Lausanne, CH-1015 Lausanne, Switzerland}
\affiliation{Department of Energy Conversion and Storage, Technical University of Denmark, Kgs. Lyngby DK-2800, Denmark}
\author{Sebastiaan\,P.\,Huber}
\email[]{mail@sphuber.net}
\affiliation{Theory and Simulation of Materials (THEOS), and National Centre for Computational Design and Discovery of Novel Materials (MARVEL), \'{E}cole Polytechnique F\'{e}d\'{e}rale de Lausanne, CH-1015 Lausanne, Switzerland}
\author{Jusong\,Yu}
\affiliation{Department of Physics, South China University of Technology, Guangzhou 510640, China}
\author{Nicola\,Marzari}
\author{Giovanni\,Pizzi}
\affiliation{Theory and Simulation of Materials (THEOS), and National Centre for Computational Design and Discovery of Novel Materials (MARVEL), \'{E}cole Polytechnique F\'{e}d\'{e}rale de Lausanne, CH-1015 Lausanne, Switzerland}

\date{\today}

\input{sections/abstract}

\maketitle

\section{Introduction}
\input{sections/introduction}

\section{User interface}
\input{sections/interface}

\section{Architecture}
\input{sections/architecture}

\section{Conclusions}
\input{sections/conclusions}

\section*{Acknowledgments}
\input{sections/acknowledgments}

\bibliography{template}

\end{document}

%% file: sections/abstract.tex
\begin{abstract}
Over the last two decades, the field of computational science has seen a dramatic shift towards incorporating high-throughput computation and big-data analysis as fundamental pillars of the scientific discovery process.
This has necessitated the development of tools and techniques to deal with the generation, storage and processing of large amounts of data.
In this work we present an in-depth look at the workflow engine powering AiiDA, a widely adopted, highly flexible and database-backed informatics infrastructure with an emphasis on data reproducibility.
We detail many of the design choices that were made which were informed by several important goals: the ability to scale from running on individual laptops up to high-performance supercomputers, managing jobs with runtimes spanning from fractions of a second to weeks and scaling up to thousands of jobs concurrently, and all this while maximising robustness.
In short, AiiDA aims to be a Swiss army knife for high-throughput computational science.
As well as the architecture, we outline important API design choices made to give workflow writers a great deal of liberty whilst guiding them towards writing robust and modular workflows, ultimately enabling them to encode their scientific knowledge to the benefit of the wider scientific community.
\end{abstract}

%% file: sections/introduction.tex
As developments in computational power have steadily and tremendously increased over the past few decades, so with them the field of computational science.
Digital applications have become increasingly complex and often comprise an intricate combination of scientific computational methods and data analysis.
The sequence of steps in these applications is often encoded in a workflow and the need to automate these processes has led to an increase in the number of \glspl{wms}.
A \gls{wms} provides the necessary functionality to define and subsequently execute workflows that essentially encode a sequence of data transformations\cite{Talia:2013}.
In recent years, many \glspl{wms} have been developed which have greatly simplified and streamlined the creation and analysis of data.
However, with these improvements come new challenges in managing the massive amounts of data that are produced.

The most apparent challenge is finding an efficient method of storing the data itself.
Although simple storage approaches can solve the problem of data persistence, they fail to address the question of data reproducibility, which carries particular importance in the scientific method and plays a direct role in making data reusable, according to the FAIR Data Principles~\cite{Wilkinson:2016}.
Indeed, the reproducibility of data can only be guaranteed if the provenance of data is treated with the same importance as the data itself, as it is the data's provenance that enables its validation and verification\cite{Ioannidis:2009,Peng:2011,Stoddart:2016,Allison:2016}.
Here, it is critical to realise that not only the data themselves, but also the workflows that create them should be part of the tracked provenance.
As a consequence of this observation, efforts are underway to extend the FAIR Data Principles to workflows as well~\cite{Goble:2020}.
With the rate of data production made possible by modern technologies, it has become untenable to reconstruct the provenance of data \emph{a posteriori}, calling for tools that automatically record it as it is created.
Some \glspl{wms} have started addressing this challenge, however, their data and workflow provenance guarantees are often insufficient to be able to retrace the origins of a piece of data or to recreate it.

In this paper, we describe in detail the workflow system of AiiDA\cite{Huber:2020}, an open-source, high-throughput, scalable computational infrastructure for automated reproducible workflows and data provenance, implemented in Python.
While the design of AiiDA and its workflow system is generic enough to be applicable to any computational, and potentially experimental, scientific domain, its origin and strengths lie in applications that make use of \gls{hpc} systems.
Users of these environments are often used to scripting, which is why the workflow engine of AiiDA provides a rich \gls{api}, unlike the majority of \gls{wms} that provide a \gls{gui}, such as Kepler\cite{Altintas}, Taverna \cite{Oinn:2004} and Triana \cite{Taylor:2003}.
An \gls{api} provides a more direct and seamless integration of the workflow system with the simulation codes and data analysis tools that it manages and that are typically used on \gls{hpc} systems.

An additional benefit of an API-based workflow language and engine is that it allows the definition of \emph{dynamic} workflows, whose exact path is not pre-determined at runtime but evolves during execution based on the results of completed steps.
In AiiDA, for example, the code that defines a workflow is directly executed by the engine and there is no intermediate translation layer.
The majority of WMSs, however, interpret workflows that are defined through a static markup language such as XML in the case of Karajan\cite{vonLaszewski:2007}, custom XML derivatives such as Askalon's\cite{Fahringer:2005} AGWL\cite{FahringerHainzer:2005} or workflow specific standards such as the \gls{cwl}\cite{Amstutz:2016}.
Some of them may provide bindings to programming languages in order to define workflows through an API, such as Pegasus \cite{Deelman:2005}, however, this is a mere pre-processing step as the workflows are still converted into a \gls{dag} representation in XML, before being executed by the workflow runner when launched.
The big disadvantage of these document based workflow definitions is that they are \emph{static}, in the sense that the exact flow of the workflow needs to be known before it is executed \footnote{Logic and loops can often be used but these need to be represented explicitly which quickly becomes cumbersome.}.
The mechanism also naturally limits the available programming structures to \glspl{dag} or directed cyclic graphs (DCG), if loops are supported by the markup language.

The recent Python-based workflow managers Signac~\cite{Adorf:2018} and Parsl~\cite{Babuji:2019} have chosen a different approach and rely on implicit dataflow to define and control workflows.
In this model, new data operations, bound by data dependencies, are executed as those dependencies are fullfilled by other data becoming available in the workspace.
The Fireworks system\cite{Jain:2015}, which supports the definition of workflows through documents in JavaScript Object Notation\cite{Bray:2014}, has made important steps toward supporting \emph{dynamic} workflows.
A workflow can insert new steps or spawn additional logical branches while it is running, based on intermediate results produced by previously completed steps.
However, while this enables runtime-mutable workflows, specific mutations are bound by the constraints of the custom static JSON markup language through which they are defined.
In contrast, workflows in AiiDA are implemented directly in Python and as such have all the dynamic expressiveness of a programming language directly at their disposal, as well as full access to the entire provenance graph with the data that is already stored in the database.
This proves to be a very powerful mechanism to deal with, for example, the problem of error handling when running high-throughput simulations.

In the field of materials science specifically, other libraries and frameworks exist that provide advanced workflows with automated handling and recovery of errors encountered in \emph{ab-initio} calculations, such as {\sc{Aflow}}\cite{Curtarolo:2012}, Atomate\cite{Mathew:2017}, MAST\cite{Mayeshiba:2017} and OQMD\cite{Saal:2013}.
However, all of these are typically only compatible with one specific \emph{ab-initio} code (with some expanding support to others), whereas the ecosystem of density functional theory (DFT) features a great variety of popular codes\cite{Lejaeghere:2016}.
By tightly coupling the \gls{wms} and the workflow implementations themselves to any single or a few codes, interoperability is naturally hamstrung.
In stark contrast, the workflow system of AiiDA is completely agnostic of the external software that performs the computation and provides an integrated abstract interface for any simulation code, with built-in support for all major resource managing systems, such as PBSPro\cite{pbspro}, SLURM\cite{slurm}, SGE\cite{sge} and Torque\cite{torque}.
Through its flexible plugin system, AiiDA allows any code to be made compatible via plugins, which are registered on the AiiDA registry\cite{aiida_plugin_registy} and can be installed with a single command through the Python package manager \texttt{pip}\cite{pip}.

With these considerations, the workflow system of AiiDA has been designed to satisfy the following criteria.
The workflow system should (i) facilitate the definition of fully \emph{dynamic} workflows, (ii) with an interface generic enough to support running arbitrary external codes, (iii) automatically store the full provenance of executed workflows (iv) in a way that makes the data easily queryable while scaling towards exascale applications (v) with an overhead of the provenance storage that does not outweigh the cost of the computational workflows themselves.
AiiDA's workflow system can be roughly split into two components: the user interface (or API) that allows the users to implement workflows and interact with the provenance graph, and the engine that is responsible for automatically executing those workflows and storing the results.
In this paper, we first describe the user interface followed by a technical description of the architecture and implementation of the engine.

%% file: sections/interface.tex
One of the main design goals of AiiDA's workflow engine interface is to minimize the restrictions imposed on the developer, while simultaneously providing the tools that enable and stimulate the development of maintainable, self-documenting, robust and modular workflows.
When forcing interaction with the process engine through a specific API, the amount of functionality disposable to the user is intrinsically limited.
To limit these restrictions to the bare minimum, workflows in AiiDA are written directly in Python and the written code is directly executed by the engine, without translational layer.
In this way, the workflow developer has direct access to the entire AiiDA API and all Python libraries, and is not dependent on a particular feature being exposed through a workflow-specific API.
Moreover, in the field of computational science, Python is an abundant, thriving and well-supported language, which means that a lot of existing code will naturally interface with AiiDA's workflow engine without any additional custom development required.

On the other hand, not restricting the methods of workflow implementation could lead to a wide variety of solutions, which almost inevitably render them incompatible and non-modular.
To counteract this undesired phenomenon while maintaining full access to the AiiDA API, the workflow engine exposes various tools and constructs to simplify the development of workflows and automatically improve robustness and interoperability.
In this section, these constructs will be explained in detail, covering their implementation and interface.

A final crucial consideration relates to the extent to which the maintenance of full data provenance, being the core principle of AiiDA, can be guaranteed by its engine.
In giving the user almost unrestricted freedom in designing and developing workflows, perfect provenance cannot be guaranteed.
As a compromise, the conditions under which provenance \emph{will} be guaranteed by AiiDA, and conversely, how it will definitely be broken, need to be as simple and clear as possible to the user.
The design mantra here is once more to restrict the user as little as possible and allow the breaking of data provenance if the user deems it necessary or justified.

\subsection{Process specification}
\label{sec:process_spec}
Any entity that can be run by AiiDA's engine is named a process and should be implemented through the class \texttt{Process}.
A process is defined as a set of logical instructions, implemented in code, that operates on a set of inputs in order to produce certain outputs, with the possibility of premature termination through known failure modes.
Each \texttt{Process} defines its inputs, outputs and known failure modes through its specification, which is facilitated by the \texttt{ProcessSpec} class.

\subsubsection{Ports and port namespaces}
\label{sec:process_spec_port_namespaces}
Before diving into the details of how inputs and outputs are specified for a process through its process specification, we clarify the concept of ports and port namespaces.
The inputs and outputs of a process share the common feature of being the gateways, or ports, through which data is ported in and out of the black box of a process.
These ports can be further grouped or nested in port namespaces.
The concepts of a port and a port namespace in AiiDA are implemented by the \texttt{Port} and \texttt{PortNamespace} classes, respectively.
The \texttt{PortNamespace} is simply a container of \texttt{Port} instances, and given that it is itself a subclass of \texttt{Port}, \texttt{PortNamespace} instances can be nested within one another.
Since the \texttt{PortNamespace} is implemented as a mapping, inserting and addressing members of the container is achieved through key referencing as with any other mapping in Python.
Each \texttt{Port} instance has the following attributes:

\begin{itemize}
    \item \texttt{valid\_type}: a tuple of accepted port value types,
    \item \texttt{validator}: a custom validator function to validate the value passed to the port,
    \item \texttt{default}: an optional default port value,
    \item \texttt{required}: a boolean to indicate whether a port value is required, and,
    \item \texttt{non\_db}: a boolean to indicate whether the port requires a database storable type.
\end{itemize}

In addition to these attributes, the \texttt{PortNamespace} has the \texttt{dynamic} attribute, which is a boolean to indicate whether the namespace can accept any values for ports that are not explicitly defined.
The use case for this concept will become clear in a later section on workchains (see Sec.\ref{sec:workchains}).
When a \texttt{Port} or \texttt{PortNamespace} is validated, the validation of each port is called recursively, which includes verifying the type of the values passed with respect to the \texttt{valid\_type} attribute and calling the \texttt{validator} function, if defined.
A \texttt{PortNamespace} is considered valid if and only if all of the ports nested within it, as well as itself, pass validation.

By default, any input to a process should be a database storable type, as otherwise the provenance of the outputs generated by that process would be lost, violating a core principle of AiiDA.
However, there are use cases where this isolated loss of provenance is acceptable and putting an absolute requirement on the storability of all inputs might be too restrictive.
For this reason, the \texttt{non\_db} attribute can be used to mark a port as not storable in the database.
That is to say, any input that is passed through a port with \texttt{non\_db} set to \texttt{True} will not be stored and linked as an input to the process node that is automatically created in the provenance graph when AiiDA executes the process.
AiiDA makes use of this feature to allow defining various process metadata, such as a label or description, which are then not stored as actual input nodes, but directly as attributes of the process node.
A user may also decide to use this feature if they deem a particular input as irrelevant for the provenance.

\subsubsection{Inputs and outputs}
\label{sec:process_spec_inputs_outputs}
The inputs and outputs of a process are defined through its specification, as implemented by the \texttt{ProcessSpec} class.
The \texttt{ProcessSpec} class contains two \texttt{PortNamespace} instances, accessible through the \texttt{inputs} and \texttt{outputs} attributes, that contain the input and output ports of the process, respectively.
To define a new input or output port for a process, the \texttt{ProcessSpec} exposes two convenience methods:

\begin{lstlisting}[
    label="code:process_spec_input_output",
    numbers=left,
    caption={The definition of an input and output port through the process specification.}
]
spec = ProcessSpec()
spec.input('a', valid_type=Int, default=Int(2), validator=is_positive_integer, required=True)
spec.output('b', valid_type=Float, default=Float(-2.0), validator=is_negative_float, required=True)
\end{lstlisting}

These two method calls result in the creation of an \texttt{InputPort}, stored in the \texttt{inputs} namespace under the key \texttt{a}, and an \texttt{OutputPort}, stored in the \texttt{outputs} namespace under the key \texttt{b}, respectively.
New input and output namespaces can be created similarly with the \texttt{input\_namespace} and \texttt{output\_namespace} methods, respectively, e.g.:

\begin{lstlisting}[
    label=code:process_spec_input_output_namespace,
    numbers=left,
    caption={The definition of an input and output port namespace through the process specification.},
]
spec = ProcessSpec()
spec.input_namespace('nested.input.namespace')
spec.output_namespace('some.outputs')
\end{lstlisting}

The argument passed to the methods is used as the key under which the newly created namespaces is inserted into their respective parent namespace.
The period is treated as a special character and is interpreted as a namespace separator.
The key \texttt{nested.input.namespace} is therefore interpreted as a nested namespace of depth three and the port namespaces are recursively created by the \texttt{input\_namespace} call.

Note that all these \texttt{ProcessSpec} methods are declarative in nature and that they can overwrite the effects of previously executed methods.
Consider the following example:

\begin{lstlisting}[
    label=code:process_spec_input_override,
    numbers=left,
    caption={The declarative nature of the process specification allows later declarations to override earlier ones.},
]
spec = ProcessSpec()
spec.input('a', valid_type=Int, default=Int(2), validator=is_positive_integer, required=True)
spec.input('a', valid_type=Float, default=Float(3.0), validator=is_positive_float, required=False)
\end{lstlisting}

The resulting process specification has a single input port \textit{a} in its inputs namespace that accepts float types, as the preceding directive is overwritten.

\subsubsection{Exit codes}
\label{sec:process_spec_exit_codes}
In addition to inputs and outputs, the process specification is also used to declare the known failure modes of the process.
A common method of communicating a particular failure from a process to its caller is through the use of an exit status.
An exit status, modelled on a similar concept found in POSIX processes, is defined as an integer that is returned by all processes. If the integer is zero, it signifies that the process executed correctly; any non-zero value indicates an error that maps onto a known failure mode.
This concept is implemented in AiiDA through `exit codes'.
Specifically, the \texttt{ProcessSpec} class implements the \texttt{exit\_code} method that allows one to define an exit code for the corresponding process:

\begin{lstlisting}[
    label=code:process_spec_exit_code,
    numbers=left,
    caption={The definition of an exit code, consisting of an integer exit status, a reference label and an exit message, through the process specification.},
]
spec = ProcessSpec()
spec.exit_code(418, 'ERROR_I_AM_A_TEAPOT', 'the process experienced an identity crisis')
\end{lstlisting}

This example defines an exit code for the process, with exit status $418$ and exit message `the workchain experienced an identity crisis'.
These two values are stored in the DB as attributes of the process.
In addition, the string \texttt{ERROR\_I\_AM\_A\_TEAPOT} is a human-readable unique label that can be conveniently used to reference the exit code instead of using the integer status.
A detailed explanation of how exit codes are referenced and used in practice is given in \cref{sec:process_spec_aborting}.

\subsection{Process implementations}

\subsubsection{Calculation functions}
\label{sec:processes_calculation_function}
To honor the design goal of restricting a workflow developer as little as possible, a solution was sought to turn any regular Python function into a fully AiiDA compliant function.
Employing the concept of Python's function decorators, a wrapping function that alters or adds to the behavior of the function it is applied to, the \texttt{calcfunction} decorator was developed.
To explain its functionality, consider the following Python functions that add and multiply two numbers, respectively:
\begin{lstlisting}[
    label=code:calcfunction_plain,
    numbers=left,
    caption={Two standard Python functions to add and multiply two numbers, respectively.},
]
def add(a, b):
    return a + b

def multiply(a, b):
    return a * b
\end{lstlisting}
By leveraging the \texttt{calcfunction} decorator, these plain Python functions are turned into AiiDA compliant functions with the addition of just a single line:
\begin{lstlisting}[
    label=code:calcfunction_decorated,
    numbers=left,
    caption={By decorating the Python function with the \texttt{calcfunction} decorator, the plain function is automatically transformed by the engine into an AiiDA process when executed.},
]
@calcfunction
def add(a, b):
    return a + b

@calcfunction
def multiply(a, b):
    return a * b
\end{lstlisting}
When either function is called, the decorator instructs the engine to create a \texttt{Process} instance on the fly, representing the function.
Python's standard \texttt{inspect} module is used to introspect the function's signature, which is used to define the input ports for the process specification as explained in Sec.\ref{sec:process_spec_inputs_outputs}.
Note that, since the process specification is generated based on the function signature, not all the functionality of ports and port namespaces are accessible.
For example, due to Python's dynamic typing, the expected type for function arguments is not always specified and therefore the \texttt{valid\_type} attributes for the generated input ports can be undefined.
We are currently working to introduce parsing of type annotations to further augment the specification of the generated \texttt{Process} to be able to type check the passed parameters.

\begin{figure}
    \centering
    \includegraphics[width=1.0\columnwidth]{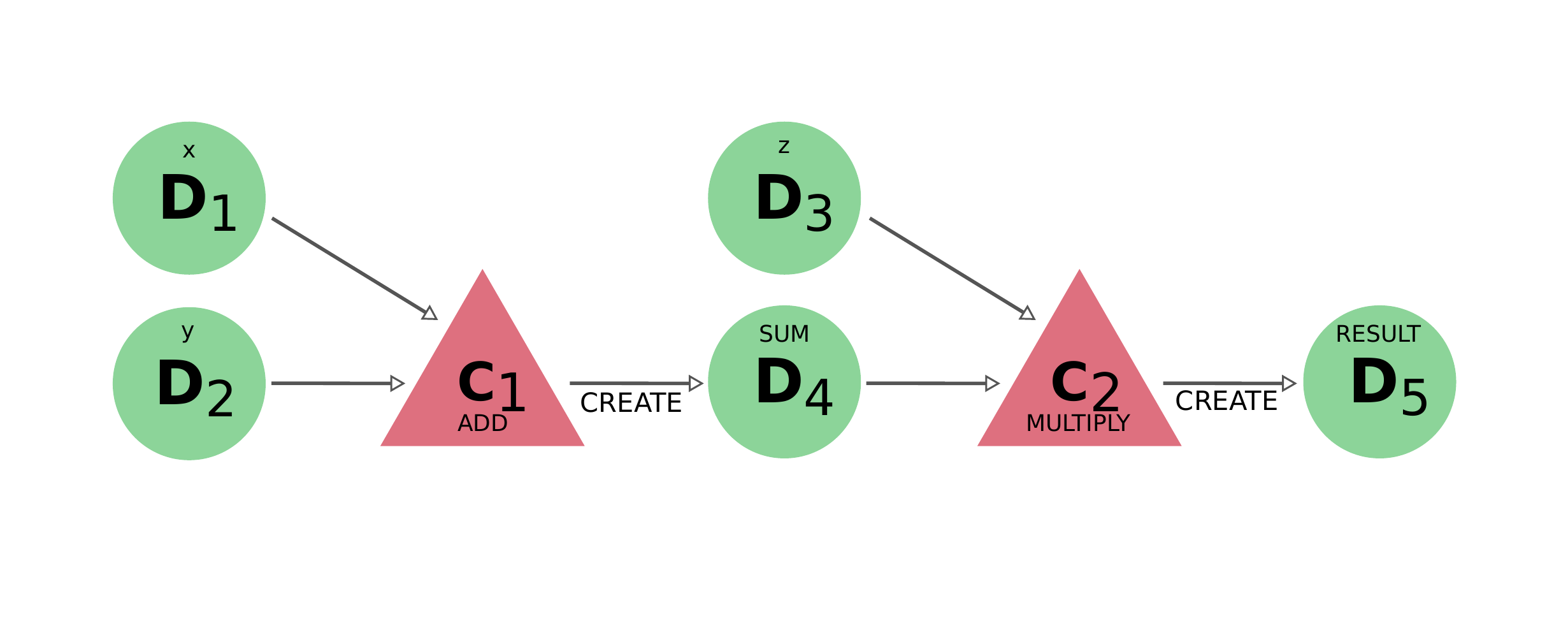}
    \caption{The provenance that is automatically generated by executing the two calculation functions.}
    \label{fig:provenance_calcfunction}
\end{figure}

By simply calling the functions with database-storable types, the engine automatically takes care of creating the corresponding data provenance in the database.
For example, the following execution:
\begin{lstlisting}[
    label=code:calcfunction_run,
    numbers=left,
    caption={Running a decorated function works just as running a normal Python function, with the only difference being that the input values should be database-storable types.},
]
multiply(add(Int(3), Int(4)), Int(5))
\end{lstlisting}
returns the value $35$ and creates a representation of the execution in the provenance graph, as represented in Fig.\ref{fig:provenance_calcfunction}.
The execution of the two functions are each represented by a \texttt{CalcFunctionNode} in the provenance graph, with the corresponding input and output nodes correctly linked to it.

\subsubsection{Work functions}
\label{sec:processes_work_function}
Even though a workflow could be encoded by means of a concatenation and/or nesting of \texttt{calcfunction} calls, that approach does not really capture the logic of a workflow.
For instance, from the created provenance graph, it is impossible to ascertain whether the two consecutively called calculations were part of a single coordinated computation, or if the output of the first was simply used as an input to an otherwise unrelated calculation.
To define a workflow that captures this logic, the engine provides the \texttt{workfunction} decorator, which is analogous to the \texttt{calcfunction}, except its purpose is not to \emph{create} new data out of its inputs, but rather to orchestrate a composition of operations.
The example of calling two calculation functions directly, as shown in listing \ref{code:calcfunction_run}, could be rewritten as follows:

\begin{lstlisting}[
    label=code:workfunction_decorated,
    numbers=left,
    caption={By decorating the Python function with the \texttt{workfunction} decorator, the plain function is automatically transformed by the engine into an AiiDA workflow when executed and `call' links are added to the calculation functions it calls.},
]

@calcfunction
def add(a, b):
    return a + b

@calcfunction
def multiply(a, b):
    return a * b

@workfunction
def add_multiply(x, y, z):
    sum = add(x, y)
    product = multiply(sum, z)
    return product

result = add_multiply(Int(1), Int(2), Int(3))
\end{lstlisting}

\begin{figure}
    \centering
    \includegraphics[width=1.0\columnwidth]{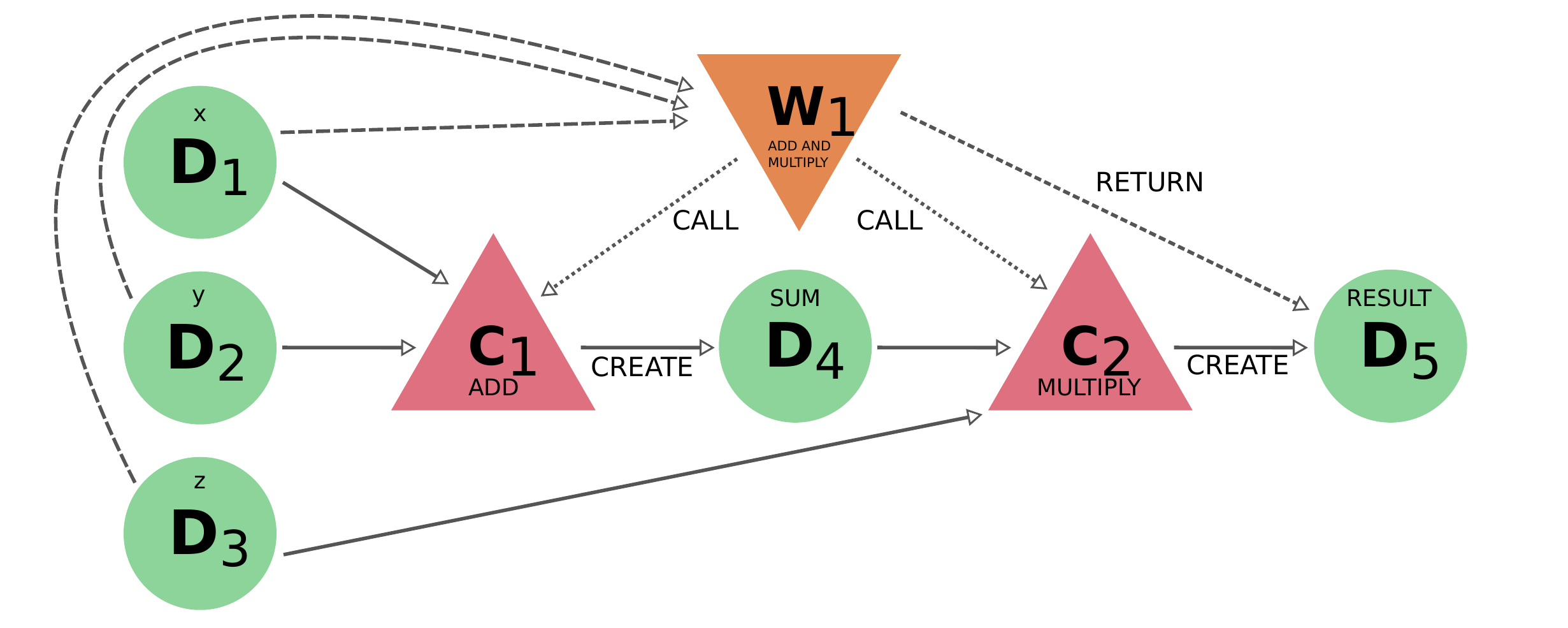}
    \caption{The provenance graph that is automatically generated by executing the a work function that calls two calculation functions in succession.}
    \label{fig:provenance_workfunction}
\end{figure}

Simply calling the decorated work function (line 16 of listing~\ref{code:workfunction_decorated}) will execute the dynamically generated process while storing a representation of it in the provenance graph as shown in Fig.\ref{fig:provenance_workfunction}.

The work function is not limited to calling only calculation functions, but can also call other work functions, and these are also linked by \texttt{CALL} links as shown in Fig.~\ref{fig:provenance_workfunction}.
Therefore, arbitrarily deeply nested workflows can be constructed with these two basic components.
However, due to their intentional simplicity, the \texttt{calcfunction} and \texttt{workfunction} decorator solutions (collectively referred to as process functions) also have inherent shortcomings that render them the incorrect tool for certain situations.

In the two examples provided earlier, the computational work that had to be performed in the function body was trivial.
However, more often than not, the opposite is the case in computational science applications.
A decorated function constitutes a contiguous block of code and will necessary block the interpreter for the duration of the function execution.
This implies that for computationally expensive functions, the interpreter will be blocked from executing anything else for extended periods of time.
Additionally, intermediate progress cannot be saved in such a way to allow execution to be resumed later at the same point in the source code.
This means that when a function is interrupted at any point, all the work it performed up to that point will be lost
(note that while Python coroutines would allow the interpreter to switch to executing other code, it would still be nearly impossible to save and later reconstruct the entire state of the call stack).
This scenario is particularly relevant considering that the simulations managed by AiiDA can last weeks, and the user might want to stop or restart the machine where AiiDA runs.

Therefore, for all its simplicity of use, process functions should be used sparingly and conscientiously in the development of workflows.
The construct implemented in AiiDA that solves all the weak points of the \texttt{workfunction} is the \texttt{WorkChain}.

\subsubsection{Work chains}
\label{sec:workchains}
The \texttt{WorkChain} class is a subclass of \texttt{Process} and is the core component of workflow development in AiiDA.
It is used to encode the logic that is typically encompassed by any scientific workflow.
The \texttt{ProcessSpec} for the \texttt{WorkChain} class supports, in addition to the standard properties of the \texttt{ProcessSpec} as detailed in section \ref{sec:process_spec}, the \texttt{outline} method.
This method allows a user to define a set of logical rules that, when evaluated, will yield a set of steps that are then executed by the engine.
The big advantage of using this construct over a concatenation of work functions is that, in between each step of a work chain, the engine automatically saves the progress to allow a restart from the last checkpoint in the case of execution failure.
Additionally, the transition between steps gives the engine the chance to yield the interpreter to other parts of the code, alleviating the blocking behavior inherent in the synchronous execution of the work function construct.

\paragraph{Outline}
\label{sec:process_spec_outline}
The \texttt{outline} method of the work chain's \texttt{ProcessSpec} is used to encode the desired logic of a certain workflow.
It supports typical logical flow constructs, such as while loops, conditional blocks and return statements.
To keep the user interface as simple as possible, the aim was to support a syntax that is as close to standard Python logical constructs as possible.
For illustration purposes consider the following description of a trivial workflow.
Print the numbers from $0$ to $100$, replacing the printed number with `fizz' if it is a multiple of three, with `buzz' if the number is a multiple of five, and with `fizzbuzz' if it is both a multiple of three and five.
The following listing demonstrates how this logic could be encoded using the syntax of the work chain outline:

\begin{lstlisting}[
    label=code:workchain_outline,
    numbers=left,
    caption={An example of a work chain outline that contains logical constructs such as a while loop and conditional statements.},
]
spec.outline(
    cls.intialize_to_zero,
    while_(cls.is_less_than_or_equal_to_hundred)(
        if_(cls.is_multiple_of_three_and_five)(
            cls.report_fizz_buzz,
        ).elif_(cls.is_multitple_of_three)(
            cls.report_fizz,
        ).elif_(cls.is_multiple_of_five)(
            cls.report_buzz,
        ).else_(
            cls.report_n,
        ),
        cls.increment_by_one,
    )
)
\end{lstlisting}

Note that, since the logical constructs \texttt{while}, \texttt{if}, \texttt{elif} and \texttt{else} are protected Python keywords, the outline analogues are suffixed with an underscore in order to properly distinguish them from the Python builtins.
The statements between or within the calls of the logical constructs represent a step that is to be executed by the engine.
Since workflow development in AiiDA happens entirely in Python, the implementation of these steps is achieved by simply implementing them as methods of the \texttt{WorkChain} class for which the outline is defined.
An example implementation of a subset of these class methods would look something like the following:

\begin{lstlisting}[
    label=code:workchain_outline_steps,
    numbers=left,
    caption={Example implementation of some of the outline steps defined in listing \ref{code:workchain_outline} as methods of the work chain class.},
]
def initialize_to_zero(self):
    self.ctx.n = 0

def is_multiple_of_three(self):
    return self.ctx.n % 3 == 0

...

def report_fizz_buzz(self):
    self.report('fizzbuzz')

...

def increment_by_one(self):
    self.ctx.n += 1
\end{lstlisting}
Each outline step is defined by a method that takes a single argument \texttt{self}, which is a reference to the class instance as is standard in Python.
This particularly trivially example already sets up the questions of how log messages can be reported from within a work chain and how data can be passed between outline steps.
The answers to these questions will be addressed in the following two sections.

\paragraph{Reporting}
\label{sec:process_spec_reporting}
The \texttt{WorkChain} class exposes the \texttt{report} method, which takes a single string as an argument, which can be used by the developer to log messages during the execution of the work chain.
The \texttt{report} method emits the passed log message through the standard Python logging module with a custom log level \texttt{REPORT}, which is defined by AiiDA and lies between the \texttt{INFO} and \texttt{WARNING} log levels.
Additionally, the logging configuration of AiiDA defines a database log handler that automatically attaches these logged messages to the node that represents the work chain from which they were emitted.
The recorded log messages can then be retrieved through the AiiDA API or its command line interface by referencing the relevant work chain node.
This report system is meant for communicating human readable log-like messages as a record of the events that occurred during its execution and should not be parsed to communicate programmatically between processes.
For that use case, the concept of exit codes has been implemented, which will be described in greater detail in \cref{sec:process_spec_aborting}.

\paragraph{Checkpoints and context}
\label{sec:process_spec_context}
As mentioned in the introduction of this section, the engine will evaluate the logic defined by the outline and execute the methods that correspond to those outline steps.
These methods only take a single argument, \texttt{self}, so there is no way to pass data directly from one step to another.
To allow data to be passed between the steps of a work chain, each \texttt{WorkChain} instance defines a context, which is a simple Python dictionary, accessible through the \texttt{ctx} attribute, that is persisted between step transitions.
The context can therefore be used as any other Python dictionary to store data and the engine ensures that the state of the work chain instance, along with its context, is saved in a checkpoint in the database (see section \ref{sec:architecture_persistence}).
Through this container, data that was stored in one outline step can be accessed in another.

\paragraph{Calling subprocesses}
\label{par:process_spec_sub_processes}
Work chains can launch any other process as a child process.
For example, one can launch a \texttt{CalcJob} (a calculation running on an external computer, described in Sec.~\ref{sec:calcjobs}) or another \texttt{WorkChain} from within a \texttt{WorkChain}.
The syntax for submitting a process from within a \texttt{WorkChain} is identical to submitting a process from a top level Python script, with the exception that one should not use the \texttt{submit} free function, but the \texttt{submit} method of the \texttt{WorkChain} class.
For example, to submit a \texttt{ChildWorkChain} with a certain set of inputs from within another work chain, one would call:
\begin{lstlisting}[
    label=code:workchain_to_context_class,
    numbers=left,
    caption={A subprocess can be submitted through the \texttt{submit} method of the \texttt{WorkChain} class and the \texttt{ToContext} container can be used to register the submitted child process as an awaitable.},
]
def submit_child_workchain(self):
    child = self.submit(ChildWorkChain, **inputs)
    return ToContext(child=child)
\end{lstlisting}
For the parent work chain to continue, it has to wait for the child process to finish and therefore it has to return control to the interpreter.
To communicate to the engine that the work chain needs to wait for a subprocess (in this example the \texttt{ChildWorkChain}) to finish, the developer should return an instance of the \texttt{ToContext} class.
This turns the submitted subprocesses into awaitables, which instructs the engine to halt execution of the work chain until all subprocesses are completed.
The same result can be achieved through the \texttt{to\_context} method of the work chain:
\begin{lstlisting}[
    label=code:workchain_to_context_method,
    numbers=left,
    caption={Similar to listing \ref{code:workchain_to_context_class}, a subprocess can be submitted through the \texttt{submit} method of the \texttt{WorkChain} class and the \texttt{to\_context} method can be used to register the submitted process as an awaitable.},
]
def submit_child_workchain(self):
    child = self.submit(ChildWorkChain, **inputs)
    self.to_context(child=child)
\end{lstlisting}
The engine will proceed to execute the submitted subprocesses and when they are completed, will assign the corresponding process node, used to represent the executed process in the database, to the specified key (\texttt{child} in this example) in the context of the parent work chain.
In the next outline step of the parent work chain, the developer will then be able to access the finished child work chain through the context member \texttt{self.ctx.child}.

One might want to submit multiple subprocesses from within a single outline step, in the case where the subprocesses are independent from one another and can be executed in parallel.
Both the \texttt{ToContext} container and the \texttt{to\_context} method do not limit the number of subprocesses that can be assigned, as long as the keys to which they are assigned are unique.
To prevent a developer from having to generate keys dynamically when one prefers to deal with an ordered list of results, the \texttt{ToContext} class and \texttt{to\_context} method both support the \texttt{append\_} free function.
Consider the following example:
\begin{lstlisting}[
    label=code:workchain_to_context_append,
    numbers=left,
    caption={The \texttt{append\_} free function can be used to alter the behavior of the \texttt{to\_context} and \texttt{ToContext} constructs to append the created awaitable to a list instead of assigning it to a specific key.},
]
def submit_multiple child_workchains(self):
    for i in range(10):
        self.to_context(children=append_(self.submit(ChildWorkChain, **inputs))
\end{lstlisting}
The \texttt{append\_} function in the \texttt{to\_context} call ensures that the subprocesses, when finished, will be appended to a list in the context under the \texttt{children} key.
In the next outline step, the developer can then access the list of process nodes that represent the completed subprocesses and iterate over them as with any other Python list.

\paragraph{Recording outputs}
\label{sec:process_spec_outputs}
To emit outputs from a work chain, the class implements the \texttt{out} method, which takes two arguments, a string (used as the outgoing link label) and a node instance.
At the point of calling, the \texttt{out} method merely records the new output node in memory and only at the end of the outline step will the engine commit the change to the database.
It is at that point that the emitted output value is validated with respect to the output port as defined in the process specification of the work chain.
When the execution of the work chain terminates, the emitted outputs are validated once more against the specification and if, for example, any required outputs have not been emitted, the work chain is marked as failed.

\paragraph{Aborting}
\label{sec:process_spec_aborting}
At any point during the execution of a work chain, a developer might want to exit from the outline logic and terminate the execution prematurely.
The engine can be instructed to terminate the execution of the work chain from within an outline step at any time, simply by returning a non-zero positive integer from the method, as shown in listing \ref{code:workchain_abort_integer}.
The non-zero positive integer return value of the outline method is interpreted as an exit status, which is set to the node that represents the work chain in the provenance graph, and the process is terminated.
\begin{lstlisting}[
    label=code:workchain_abort_integer,
    numbers=left,
    caption={By returning a non-zero positive integer from any outline method, the engine is instructed to terminate the execution of the work chain and the return value is set as the \texttt{exit\_status} attribute on the corresponding node.},
]
def abort_from_this_step(self):
    self.report('work chain will be terminated')
    return 404
\end{lstlisting}
Alternatively, to provide an accompanying message for the reason of the exit, an instance of the \texttt{ExitCode} named tuple can be returned as well, which has the same effect as the integer exit status.
The named tuple consists of an integer exit status and a string exit message.
The tuple can be constructed manually, or it can be retrieved through the \texttt{exit\_codes} attribute of the work chain, which is a container of the exit codes defined through the process specification of the work chain, as shown in listing \ref{code:process_spec_exit_code}.
\begin{lstlisting}[
    label=code:workchain_abort_exit_code,
    numbers=left,
    caption={By returning a \texttt{ExitCode} named tuple instance, the engine is instructed to terminate the execution of the work chain and the \texttt{exit\_status} and \texttt{exit\_message} of the return value is set on the corresponding node.},
]
class AbortingWorkChain(WorkChain):

    @classmethod
    def define(cls, spec):
        super().define(spec)
        spec.exit_code(404, INEVITABLE_ERROR, 'this was unavoidable')
        spec.outline(cls.abort_straightaway)

    def abort_straightaway(self):
        self.report('work chain will be terminated')
        return self.exit_codes.INEVITABLE_ERROR
\end{lstlisting}
The \texttt{self.exit\_codes.INEVITABLE\_ERROR} call retrieves the exit code instance that was defined in the process specification (line 6) and, when returned from the outline step, triggers the engine to terminate the work chain.
The exit status and exit message of the exit code is set to the corresponding attributes of the work chain node.
Any potential caller of the work chain can then inspect these attributes and, based on their value, decide how to proceed.

\paragraph{Exposing of ports}
\label{sec:expose_ports}
As mentioned in the introduction, one of the major design goals of the workflow environment in AiiDA is to limit developers as little as possible in their freedom to design solutions, while providing them with the tools to write modular workflows.
Modular workflows in this sense can be defined as workflows that perform a single well-defined task.
Higher level workflows can then easily be built by wrapping these lower-level blocks.
When a work chain wraps another work chain, it needs to `expose' its input (and potentially output) ports, such that the caller of the top level work chain can pass in the required inputs.
To make the process of wrapping a work chain within another workchain as simple as possible, and to prevent a developer from having to copy the port specification of the wrapped work chain manually, AiiDA implements the concept of automatic port exposing.
To illustrate the concept of port exposing, consider the simple example of a \texttt{ParentWorkChain} wrapping a \texttt{ChildWorkChain}.
\begin{lstlisting}[
    label=code:workchain_expose_inputs,
    numbers=left,
    caption={The \texttt{expose\_inputs} method of the \texttt{ProcessSpec} class allows a work chain to automatically copy the ports of the work chain it is wrapping.},
]
class ParentWorkChain(WorkChain):

    @classmethod
    def define(cls, spec):
        super().define(spec)
        spec.expose_inputs(ChildWorkChain)
        spec.outline(cls.run_child)

    def run_child(self):
        child_inputs = self.exposed_inputs(ChildWorkChain)
        child = self.submit(ChildWorkChain, **child_inputs)
        return ToContext(child=child)

class ChildWorkChain(WorkChain):

    @classmethod
    def define(cls, spec):
        super().define(spec)
        spec.input('a', valid_type=Int)
        spec.outline(cls.run_step)

    def run_step(self):
        self.report('running the ChildWorkChain')
\end{lstlisting}

The \texttt{expose\_inputs} method by default copies over all the ports of the child work chain.
Optionally, ports can be omitted through the \texttt{exclude} keyword, or specific ports can be selected with the \texttt{include} keyword.
The work chain ports can also be exposed in a particular namespace by using the \texttt{namespace} keyword.
This is especially useful if the exposed ports would otherwise overlap with existing ports with the same name.

\subsubsection{Calculation jobs}
\label{sec:calcjobs}
High-performance computing resources rarely allow their users to directly run calculations on the system, but instead require resources to be requested from, or jobs to be submitted to, a scheduler, such as \href{https://hpc.llnl.gov/tag/software/slurm}{SLURM} or \href{http://www.arc.ox.ac.uk/content/pbs}{PBS}.
Submitting calculations as jobs to these schedulers on remote computing resources is one of the most common activities in the workflow of a computational scientist, but involves a substantial amount of repetitive work.
The job script has to be prepared and uploaded to the remote machine, including any other required input files for the calculation that is to be run.
Subsequently, the job has to be submitted to the scheduler which will put it in a queue.
The user must then monitor the queue to determine when the job is completed and the output files can be retrieved from the remote computing resource, to be optionally parsed and passed through post-processing tools.

This entire process is automated by AiiDA and implemented by the \texttt{CalcJob} class.
A detailed description of how the \texttt{CalcJob} can be implemented for an arbitrary code that can be run on a remote cluster is beyond the scope of this paper and can be found in the extensive online documentation~(\href{https://aiida-core.readthedocs.io/en/latest/}{aiida-core.readthedocs.io}).
Here, rather, we focus only on how the required steps of running a calculation job are realized automatically by AiiDA's engine.

In addition to having to run a calculation through a scheduler, another complexity of running calculations on high-performance computing resources is that these machines have to be remotely accessed, typically over an SSH connection.
Since AiiDA is not required to run on the computing resource itself, it needs the ability to open a connection to perform the various operations involved with running calculation jobs.
Any operation that requires opening a connection to the remote computing resource is referred to as a `transport task', as it requires the SSH connection to ``transport'' the command from the local machine where AiiDA is running to the remote machine.

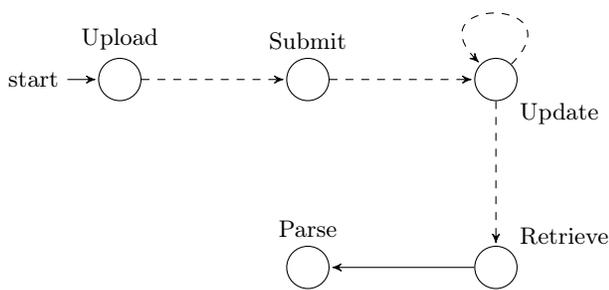
\begin{figure}
\center
\begin{tikzpicture}[>=stealth',shorten >=1pt,auto]
    \tikzstyle{every state}=[node distance=2.5cm, font=\tiny, minimum size=16pt]

    \node[state,initial,label=above:{Upload}] (uploading) {};
    \node[state,right of=uploading, label=above:{Submit}] (submitting) {};
    \node[state,right of=submitting, label=below right:{Update}] (updating) {};
    \node[state,below of=updating, label=above right:{Retrieve}] (retrieving) {};
    \node[state,left of=retrieving, label=above:{Parse}] (parsing) {};

    \path[->] (uploading) edge[dashed] (submitting)
              (submitting) edge[dashed] (updating)
              (updating) edge [loop,dashed] (updating)
              (updating) edge[dashed] (retrieving)
              (retrieving) edge (parsing);

%     \path[->,dashed] ($(parsing) + (-3,0)$) edge node[anchor=north] (w-transport){$=$} +(1,0) ;
%     \node[anchor=north] at (w-transport) {With transport};

\end{tikzpicture}
\caption{
    Substates of the calculation job (see \cref{fig:process_state_machine} for the full process state machine).
    A remote job is executed by transitioning through these states with each dashed transition occurring only when an open transport is available (prior to which the process is in a waiting state which is not shown).
}
\label{fig:calcjob-state-machine}
\end{figure}

The life cycle of each calculation job knows four transport tasks that are executed in succession as shown in \cref{fig:calcjob-state-machine}.
For the first task, `upload', the engine creates a new folder in the scratch space on the remote machine, into which the input files and job script are uploaded.
Subsequently, the `submit' task executes the command to submit the newly uploaded job script to the scheduler.
If the job is successfully submitted, the output of the command is parsed to retrieve the unique identifier that the scheduler assigned to the job.
The engine then uses this identifier to query the status of the job calculation in the `update' task.
Once the scheduler returns that the job has been completed, the engine invokes the `retrieve' task, which retrieves all the files specified by the calculation plugin from the remote working directory to a local folder.
This folder is attached as an output to the calculation node that represents the execution of the calculation job in the provenance graph.
Finally, the engine parses the retrieved data and attaches the resulting output nodes to the formerly mentioned calculation node.
However, since the data is already retrieved to the local machine, this operation does not require an open SSH connection and is therefore not a transport task.

\paragraph{Error handling and robustness}
As detailed in the previous section, the majority of operations required for running a calculation job to completion on a remote computing resource require the opening of an SSH connection and executing a command over that connection.
A variety of problems may occur during these steps.
For example, the remote machine may be unreachable because the client machine lost its network connection, or the remote machine itself has network issues.
But even when a connection is successfully established, there are still a myriad of reasons why the remote operations may fail.
The operation can be interrupted, timeout or simply fail on the remote machine.
The latter can occur often for the transport tasks that interact with the scheduler, such as the `submit' and `update' tasks, when the scheduler is unreachable or overloaded.

In any case, problems like these during the execution of transport tasks raise exceptions that, however, should not cause a failure of the running process but rather be dealt with elegantly.
More importantly, the character of these problems are often temporary and often resolve themselves if not with a little intervention from the user.
These transport tasks therefore benefit enormously from an automated retry mechanism that detects problems and automatically retries the operation at a later point in time.
This concept has been implemented in AiiDA's engine by an `exponential-back-off-retry' mechanism, that works as follows.
Each transport task is wrapped in an exponential-back-off-retry co-routine.
This wrapper catches any exceptions that occur during the execution of the transport task, in which case it reschedules the same operation for execution at a later point in time.
The wrapper reschedules failed tasks a maximum number of times, each time doubling the initial waiting interval, after which the parent process to which the transport task belongs, is paused.
The initial waiting interval and the maximum number of retries are configurable per type of transport task.

In our experience, many problems are automatically resolved by this exponential-back-off-retry mechanism, but if this is not the case, the engine pauses the process instead of letting it except, giving the user the opportunity to investigate the problem.
If the cause for the exception was external to the process itself and can be fixed, the paused process can be successfully resumed by the user.
Alternatively, if the source of the exception was with the process itself (e.g., a coding error in a parser or in a workflow step), the user can manually kill it.

This mechanism of making calculation job processes robust is indispensable in high-throughput workflows.
Without it, if, for instance, the network connection is lost when running a large number of complex and nested workflows in parallel, then all of the calculation jobs would except and, in a big cascade, all their parent processes too, resulting in a substantial loss of work.
The exponential-back-off-retry mechanism combines a high degree of automation, by autonomously retrying failed tasks, and ultimately pausing tasks with problems instead of letting them except.

\paragraph{Transport queue}
With the engine being primarily designed to operate under high-throughput conditions, one often runs many calculation jobs on a given remote resource, which requires many connections to be opened to that machine.
However, remote computing resources often limit the amount of connections that are allowed to be opened in a given time interval by a single client.
Exceeding said limit can lead the client to be banned entirely from accessing the machine.
To reduce the number of opened connections, while maintaining high-throughput capability, the engine bundles all connections through a `transport queue'.
Each worker (an independent Python process executing workflows - see \cref{sec:daemon} for details) maintains one transport queue and the calculation jobs that it manages make requests for an open transport, instead of opening one themselves whenever they need it.
The worker collects these requests and, at given point in time, opens a single connection to the remote machine and distributes the transport to the processes that requested it.
The transport queue guarantees that it opens a connection only once per safe-interval, a configurable minimum amount of time allowed between connections.
This mechanism ensures that the maximum connection opening rate is never exceeded, even when running many concurrent calculation jobs on the same machine.
The only limitation of this mechanism is that each worker maintains its own transport queue and there is no communication between those queues.
This means that the promise of the safe-interval is only guaranteed per worker.
However, by knowing how many workers are active (a value that the user can decide) the interval can of course be configured such that, on average, the connection rate is respected across all active workers.

\paragraph{Bundling scheduler update requests}
The bundling of connections required by calculation jobs through the transport queue, as described in the previous paragraph, already relieves most of the load on remote computing resources when running in high-throughput mode.
However, each active calculation job would still regularly perform the required remote operations, such as querying the scheduler for the state of the job, separately.
When running in high-throughput mode, this scheme can still put unacceptable loads on the scheduler, despite the connection being shared.
This, in turn, can render the scheduler unusable for all users of the computer cluster.
Therefore, the AiiDA engine bundles all scheduler updates for calculation jobs (on each worker), very similarly to the transport queue for connections.
When a calculation job needs to update its status, instead of polling the scheduler directly, it schedules an update request with the job manager of the worker.
The job manager records these requests and, once a remote connection becomes available from the transport queue, issues a single scheduler update for the job identifiers that have registered themselves with it.
The response is then parsed and the new status of each registered job is communicated to the corresponding calculation job.
The combination of the bundling of connections and scheduler update requests ensures that the engine can run concurrent calculations jobs without overloading the remote computing resource.

%% file: sections/architecture.tex
\begin{figure}
    \input{images/db-rmq.tex}
    \caption{One or more clients and one or more workers maintain a connection via TCP/IP to the database and RabbitMQ service enabling a rich set of possible configurations and corresponding usage scenarios.}
    \label{fig:client-worker}
\end{figure}
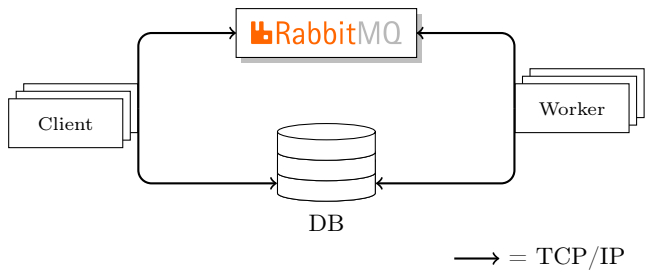

AiiDA's software architecture reflects several design goals that are informed, principally, by the needs of the high-throughput materials science community.
These include the ability scale from running on laptops up to high-performance supercomputers, carrying out processes that range anywhere from fractions of a second to, potentially, weeks in execution time.
Furthermore, it should be possible to have up to thousands of processes simultaneously active in a single instance.
In terms of deployment configurations, AiiDA instances are typically installed on each user's workstation.
However, there is the possibility to have a group or organisation-wide deployment or even a public-serving instance, as employed by the Materials Cloud \citep{Talirz2020}.

As shown in \cref{fig:client-worker}, the workflow engine relies on two main external components for the execution of workflows:
\begin{enumerate}[label={\alph*)}]
    \item The database engine (PostgreSQL\footnote{\url{https://www.postgresql.org/}}), used to persist the state of currently running processes, which doubles as a proxy to reflect the state of processes to the user, and,
    \item the message broker (RabbitMQ\footnote{\url{https://www.rabbitmq.com/}}), which is responsible for delivering messages between the client(s) and the worker(s) (which may run in the same Python instance or even be on separate computers).
\end{enumerate}
This decoupled approach has numerous advantages both in terms of flexibility of deployment configurations and for enabling a clear separation of concerns that makes it easier to write correct and robust code.

\subsection{The engine}
In order to meet a number of the design goals for the workflow system relating to responsiveness, scaling and high-throughput capability, we rely heavily on events to trigger actions such as progressing a workflow from a state where it was waiting on something to complete, or to initiate the orderly termination of a running workflow.
This is in stark contrast to polling based systems, where an entity that is waiting for a particular outcome has to periodically check if the event has occurred, often leading to unnecessary loads on the database and to poor responsiveness.

As with many event-based systems (e.g. graphical user interfaces, computer games) AiiDA uses an event loop to achieve this, which in our case is provided by Python's built-in \texttt{asyncio}.
Furthermore, this enables multiple AiiDA processes to be managed by a single Python instance, despite the lack of multithreading support, as coroutines can be scheduled for execution and can yield to others when they are waiting for some action to complete.
The use of coroutines mitigates another issue, which is that database servers typically have a low default connection limit (100 in the case of PostgreSQL) and in threaded environments it is not uncommon for each thread to have a separate connection.  Lastly, writing correct multithreaded code is extremely complex and would be difficult, even for experienced programmers, particularly given that we place no restriction on the API calls that can be made.

The entire stack of Python components needed to execute workflows are brought together in the \texttt{Runner} class which provides the event loop, persistence, communication, transport (e.g., SSH) and other functionality, some of which are described in greater detail below.
Thanks to the event loop, each runner can run any number of workflow processes concurrently (within memory limits).
We call the number processes that can run on a single runner the number of process slots.

\subsubsection{The daemon}
\label{sec:daemon}
The runner can be used in a local interpreter.
However, in most production environments the user wishes to launch a daemon that can manage one or more runners, automatically restarting them if they happen to crash.
In AiiDA, we use the Circus\cite{circus} library to achieve this.
Circus provides the ability to start multiple operating system processes, automatically restarting them if they crash.
In addition, it can show information about their current resource usage and dynamically increase the number of runners in the pool.

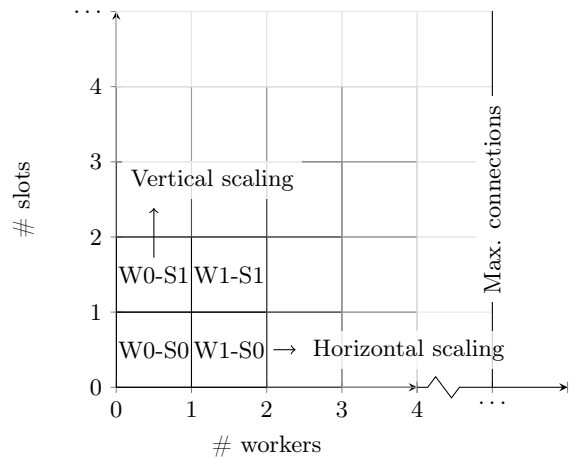
\begin{figure}
    \centering
\begin{tikzpicture}[auto]

\begin{groupplot}[
    group style={
        group size=2 by 1,
        xticklabels at=edge bottom,
        horizontal sep=0pt
    },
    axis lines = left,
    ylabel = {\# slots},
    ymin=0, ymax=5,
    xmin=0, xmax=101,
    ytick={0, 1, 2, 3, 4, 5},
    yticklabels = {0, 1, 2, 3, 4, $\ldots$},
    scale only axis,
]
\nextgroupplot[
    xmin=0,xmax=4,
    ymin=0,ymax=5,
    width=4cm,
    height=5cm,
    xlabel = {\# workers},
]
\draw (axis cs: 0,0) rectangle node[anchor=center] {W0-S0} (axis cs: 1,1);
\draw (axis cs: 0,1) rectangle node[anchor=center] (w1s2) {W0-S1} (axis cs: 1,2);
\draw[color=black!70] (axis cs: 0,2) rectangle node[anchor=center] (w1s3) {} (axis cs: 1,3);
\draw[color=black!40] (axis cs: 0,3) rectangle (axis cs: 1,4);
\draw[color=black!10] (axis cs: 0,4) rectangle (axis cs: 1,5);

\draw (axis cs: 1,0) rectangle node[anchor=center] (w2s1) {W1-S0} (axis cs: 2,1);
\draw (axis cs: 1,1) rectangle node[anchor=center] {W1-S1} (axis cs: 2,2);
\draw[color=black!70] (axis cs: 1,2) rectangle (axis cs: 2,3);
\draw[color=black!40] (axis cs: 1,3) rectangle (axis cs: 2,4);
\draw[color=black!10] (axis cs: 1,4) rectangle (axis cs: 2,5);

\draw[color=black!70] (axis cs: 2,0) rectangle node[anchor=center] (w3s1) {} (axis cs: 3,1);
\draw[color=black!70] (axis cs: 2,1) rectangle (axis cs: 3,2);
\draw[color=black!70] (axis cs: 2,2) rectangle (axis cs: 3,3);
\draw[color=black!40] (axis cs: 2,3) rectangle (axis cs: 3,4);
\draw[color=black!10] (axis cs: 2,4) rectangle (axis cs: 3,5);

\draw[color=black!40] (axis cs: 3,0) rectangle (axis cs: 4,1);
\draw[color=black!40] (axis cs: 3,1) rectangle (axis cs: 4,2);
\draw[color=black!40] (axis cs: 3,2) rectangle (axis cs: 4,3);
\draw[color=black!40] (axis cs: 3,3) rectangle (axis cs: 4,4);
\draw[color=black!10] (axis cs: 3,4) rectangle (axis cs: 4,5);

\draw[color=black!10] (axis cs: 4,0) rectangle (axis cs: 5,1);
\draw[color=black!10] (axis cs: 4,1) rectangle (axis cs: 5,2);
\draw[color=black!10] (axis cs: 4,2) rectangle (axis cs: 5,3);
\draw[color=black!10] (axis cs: 4,3) rectangle (axis cs: 5,4);
\draw[color=black!10] (axis cs: 4,4) rectangle (axis cs: 5,5);

\draw[->] (w2s1) -- (w3s1);
\draw[->] (w1s2) -- (w1s3);

\nextgroupplot[
    xmin=99, xmax=101,
    ymin=0, ymax=5,
    xtick = {99,100,101},
    xticklabels={,$\ldots$,},
    axis y line=none,
    axis x line=middle,
    axis x discontinuity=crunch,
    height=5cm,
    width=2cm
]

\draw[color=black!10] (axis cs: 99,1) rectangle (axis cs: 100,2);
\draw[color=black!10] (axis cs: 99,2) rectangle (axis cs: 100,3);
\draw[color=black!10] (axis cs: 99,3) rectangle (axis cs: 100,4);
\draw[color=black!10] (axis cs: 99,4) rectangle (axis cs: 100,5);

\draw (axis cs:100,0) -- (axis cs:100,6);
\node[rotate=90,fill=white,anchor=center] at (axis cs:100,2.5) {Max. connections};

\end{groupplot}

\node[anchor=south west, xshift=-12pt,fill=white,fill opacity=0.8,text opacity=1] at (w1s3) {\small Vertical scaling};
\node[anchor=west,fill=white,fill opacity=0.8,text opacity=1] at (w3s1) {\small Horizontal scaling};

\end{tikzpicture}
    \caption{
        Scaling of the number of workers and slots per worker.
        The total number of concurrent processes is the product of the values on the two axes.
    }
    \label{fig:daemon-scaling}
\end{figure}

With the daemon one can scale both vertically (multiple slots per runner) and horizontally (multiple Python instances, each with one runner) as shown in \cref{fig:daemon-scaling}.
For workloads that require significant in-Python processing it is preferable to scale the number of workers while workloads involving many remote calculations can just as well scale the number of slots per worker, minimising the load on the computer running the daemon with little loss in throughput.
The hard limit is reached when the number of workers equals the maximum number of database connections (however, this is user configurable if they have access to the database settings).

\subsection{The process}

\begin{figure}
\center
\begin{tikzpicture}[>=stealth',shorten >=1pt,auto]
%                   ___
%                  |   v
%     CREATED --- RUNNING --- FINISHED (o)
%                  |   ^     /
%                  v   |    /
%                  WAITING--
%                  |   ^
%                   ----

%       * -- EXCEPTED (o)
%       * -- KILLED (o)
    \tikzstyle{every state}=[node distance=2.5cm, font=\tiny, minimum size=16pt]
    \node[state,initial,label=above:{Created}] (created) {};
    \node[state,right of=created, label=above right:{Running}] (running) {};
    % Terminal states
    \node[state,accepting,right of=running, label=below:{Finished}] (finished) {};
    \node[state,accepting,below of=created, label=below:{Excepted}] (excepted) {};
    \node[state,accepting,below of=running, label=left:{Killed}] (killed) {};

    \node[state,below of=finished, label=below:{Waiting}] (waiting) {};

    \path[->] (created) edge (running)
              (running) edge (finished)
              (running) edge [loop above] (running)
              (running) edge [bend right=15] (waiting)
              (waiting) edge [bend right=15.] (running);

    \path[->] (created) edge (killed);
    \path[->] (running) edge (killed);
    \path[->] (waiting) edge (killed);

    \path[->] (created) edge (excepted);
    \path[->] (running) edge (excepted);
    \path[->] (waiting) edge[bend left] (excepted);
\end{tikzpicture}
\caption{
    The process state machine.
    Terminal states are represented with a double circle.
}
\label{fig:process_state_machine}
\end{figure}
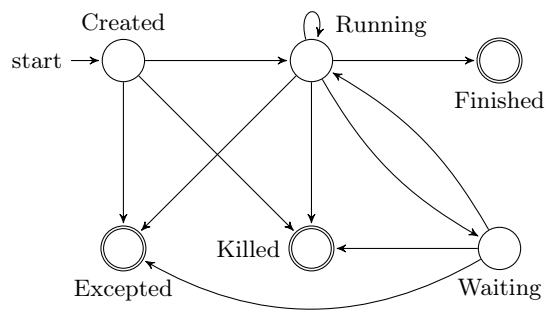

The principal object of AiiDA's workflow engine is the \texttt{Process} class.
All specific classes (\texttt{WorkChain}s, \texttt{CalcJob}s, etc.) derive from \texttt{Process} (or a subclass thereof) and in so doing inherit a large swathe of common functionality and features.
The \texttt{Process} class itself is modeled as an extended state machine, meaning that it is composed of a finite-state machine, shown in \cref{fig:process_state_machine}, where each state can have internal data members as part of its extended state.

This is a pattern common in event-driven systems as it provides a consistent way to model the current state, as well as event hooks that can be used as triggers to perform actions either internal or external to the process during state transitions.
The event hooks themselves come in the form of process member functions, e.g.:

\begin{lstlisting}[
    language=python,
    label=code:state-transition-hooks,
    caption={AiiDA's \texttt{Process} state transition hooks.
    These are invaluable for being able to guarantee that certain actions are performed when a state transition occurs.},
]
# Entering a new state
def on_entering(self, state):
    ...

# Just entered the new state from 'from_state'
def on_entered(self, from_state):
    ...

# About to exit the current state
def on_exiting(self):
    ...
\end{lstlisting}

By using these hooks it is possible to schedule actions that should always be executed at the various points of a state transition and therefore guarantee a consistent state once the \texttt{on\_entered} hook has finished.
One use of these hooks in AiiDA is to reflect the current state of the process back to the database including saving a checkpoint.
The state transition hooks are also used to send broadcast messages that allow listeners (which can potentially be on remote machines) to be updated of state changes as they happen (see \cref{sec:arch-comm}).

A typical process progresses through the various states, potentially running several member functions or waiting on other processes to finish.
However, if an exception is propagated up to the \texttt{Process} level, it will be caught and the \texttt{Process} transitions to the terminal \texttt{EXCEPTED} state, whereupon a log entry is created containing, amongst other things, the Python stack trace.
The other way to terminate a \texttt{Process} prematurely is to call the kill method.

\subsubsection{Persistence}
\label{sec:architecture_persistence}
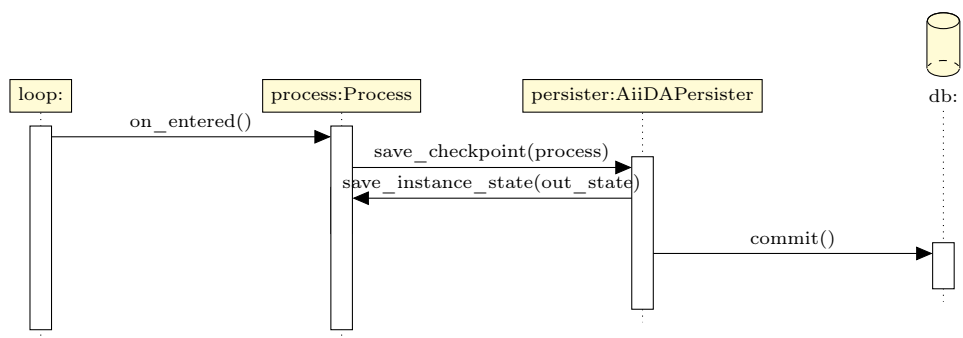
\begin{figure*}
\tikzumlset{font=\scriptsize}
\begin{tikzpicture}
\begin{umlseqdiag}
% Entities required for interaction
\umlobject[]{loop}
\umlobject[class=Process]{process}
\umlobject[class=AiiDAPersister]{persister}
\umldatabase[]{db}
% State transition
\begin{umlcall}[op=on\_entered(), type=synchron]{loop}{process}
    \begin{umlcall}[op=save\_checkpoint(process), type=synchron]{process}{persister}
        \begin{umlcall}[op=save\_instance\_state(out\_state), type=synchron]{persister}{process}\end{umlcall}
        \begin{umlcall}[op=commit(), type=synchron]{persister}{db}\end{umlcall}
    \end{umlcall}
\end{umlcall}
\end{umlseqdiag}
\end{tikzpicture}
\caption{
    A UML activity diagram showing the persistence of the internal state of a process at a state transition.  The vertical bars show the period during witch the entity (at the top) is active and therefore defines its scope.
    When the persister gets a request to save the checkpoint is calls the Process and requests that it populates a dictionary, \texttt{out\_state} which is then serialized and committed to the database.
}
\label{fig:process_persistence}
\end{figure*}

A key requirement for the engine is that the progress of processes is check-pointed and persisted such that, in the case of an orderly or disorderly shutdown, the AiiDA engine can continue from the last clean state upon being restarted.
In order to achieve this we use checkpoints that are written to the database at process state transitions.
Specifically, the context of the process, outputs, and some metadata are saved to a dictionary which is then serialized to the database by an AiiDA specific persister as shown in \cref{fig:process_persistence}.

\subsection{Communication}
\label{sec:arch-comm}

The workflow engine relies heavily on messaging for the external control of processes and to maintain high-throughput whilst ensuring fault tolerance.
This is achieved by using a ``message broker'', in our case RabbitMQ.
Message brokers typically take responsibility for guaranteeing the durability and atomicity of messages, allowing the application to focus on the business logic.
In RabbitMQ's case the user installs a service that client software interacts with via a TCP port.
The routing and persistence of messages are handled internally by RabbitMQ.

To facilitate AiiDA's interaction with RabbitMQ, we developed a software library, kiwiPy \citep{Uhrin2020}, which in turn relies on the aio\_pika \footnote{\url{https://aio-pika.readthedocs.io/en/latest/}} library.
KiwiPy significantly simplifies the process of interacting with RabbitMQ and provides the ability to offload communication to a separate thread.
This is essential for AiiDA as described below in subsection `task queues'.
In addition to task queues kiwiPy provides AiiDA the ability to send \gls{rpc} and broadcast messages.

\paragraph{Task queues} These are used to schedule new processes to be run.
There is a major advantage to using RabbitMQ in that it provides certain guarantees about messages, depending on the chosen settings.
For our task queues AiiDA uses persistent messages, which are persisted to disk such that they survive intentional or unintentional restarts of the machine.
As such, a job is guaranteed to never be lost once delivered from the client to the broker.
Furthermore, RabbitMQ expects acknowledgements for tasks that have been completed.
If it loses connection with the worker, it automatically requeues the task again, until completion is acknowledged.
This mechanism relies on the use of periodic messages, called heartbeats, to which the worker must respond in a timely manner, otherwise, upon missing two consecutive responses, RabbitMQ assumes the worker to be dead and triggers the rescheduling mechanism.
It is for this reason that kiwiPy runs a separate thread so that, even when AiiDA processes are under a heavy and blocking workload, it is able to respond to heartbeats.

\paragraph{\gls{rpc}} As suggested by the name, these kinds of messages are used to invoke a procedure (in our case a function or method) on the receiving process and deliver the result of the process back to the caller.
This is used primarily to pause, play and kill active processes.

\paragraph{Broadcast} This involves sending a single message to any registered listeners with no possibility for them to send a direct response.
These are used for two purposes: to pause, play or kill groups of processes and to control the flow between them.
Parent processes that have spawned children can choose to wait for a child to complete before continuing their execution.
This is facilitated by registering itself as a listener to broadcasts from the child and yielding until it receives the child ``terminated'' message.
This mechanism is what enables the functionality of the \texttt{to\_context} work chain construct, as described in \cref{par:process_spec_sub_processes}, notifying the work chain that it can continue as the process it was waiting for has completed.

%% file: images/db-rmq.tex
\begin{tikzpicture}[auto, transform shape]

  \newcommand\DrawDB[6]{
    % #1 = No. tiers #2 = Min width, #3 = Min height, #4 = Aspect, #5 = Position, #6 = Name of main node

    \node[draw, cylinder, shape aspect=#4,
      rotate=90, minimum width=#2, minimum height=#3](#6) at #5 {};
    \coordinate(temp_height) at ($(#6.top) - (#6.bottom) - (#6.after top) + (#6.before bottom)$);
    \newlength\heighttop
    \pgfextracty\heighttop{\pgfpointanchor{temp_height}{center}}
    \newlength\tierheight
    \pgfmathsetlength{\tierheight}{(#3 + 2*\heighttop)/3}

    \coordinate(next_cyl) at (#6.before bottom);
    \foreach \i in {1,2..,#1} {%
      \node[draw, fill=white, cylinder, shape aspect=1,
	rotate=90, minimum width=#2, minimum height=\tierheight, anchor=before bottom](cyl) at (next_cyl) {};
      \coordinate (next_cyl) at (cyl.after top);
    }
  }

  % Define block styles
  \tikzstyle{block} = [rectangle, draw, fill=blue!20,
      text width=5em, text centered, rounded corners, minimum height=4em]
  \tikzstyle{cloud} = [draw, fill=white, execute at begin node=\scriptsize, text width=4em, node distance=3cm, text centered, minimum height=2em]
  \tikzstyle{ann} = [above, text width=5em, text centered]
  \tikzstyle{texto} = [above]

  % Arrows
  \tikzstyle{single} = [draw, ->, thick, rounded corners=5pt]
  \def\blockdist{0.9cm}
  \def\toolsdist{0.9cm}

  \pgfdeclarelayer{background}
  \pgfdeclarelayer{foreground}
  \pgfsetlayers{background,main,foreground}

  % Place nodes
  \node[cloud,anchor=east] (client) {Client};
  \node[cloud,anchor=east] at (-0.1, -0.1) (client2) {Client};
  \node[cloud,anchor=east] at (-0.2, -0.2) (client3) {Client};

  \DrawDB{3}{4em}{2em}{1}{(2.5,-1)}{db}
  \node[below of=db, node distance=0.3cm, text width=2cm, text centered, anchor=north] (db_label) {DB};

  \node[draw, drop shadow, minimum height=2em, fill=white, minimum width=2.4cm, inner sep=0pt, outer sep=0pt] (rmq) at (2.5,1)  {\includegraphics[width=2cm]{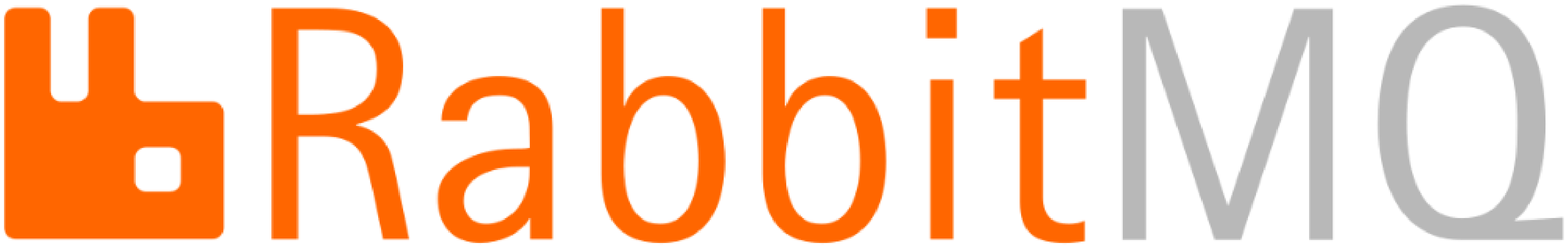}};

  \node[cloud,anchor=west] (worker) at (5.2,0.2) {Worker};
  \node[cloud,anchor=west] (worker) at (5.1,0.1) {Worker};
  \node[cloud,anchor=west] (worker) at (5,0) {Worker};

  % Draw edges
  \draw [single] (client.east) |- (rmq);
  \draw [single] (client.east) |- (db);
  \draw [single] (worker.west) |- (rmq);
  \draw [single] (worker.west) |- (db);

  \draw [single] (4.2, -2) -- (4.8, -2) node[anchor=west] {= TCP/IP};

\end{tikzpicture}

%% file: sections/conclusions.tex
In this article we have described the decisions that have guided the development and shaped the internals of AiiDA's workflow engine, in the recognition that often the insights gained in the development process can be as valuable as the finished product itself.
AiiDA has a fairly broad set of challenging goals and target use cases.
To meet these, we have incorporated a number of advanced programming techniques such as coroutines, extended finite-state machines, event-driven programming, futures, and a number of technologies commonly employed in industry, but less common in academia, such as RabbitMQ and PostgreSQL.
Being mindful of the inherent complexity of this system, and the tasks it aims to address, we have attempted to create a user-friendly API that allows non-experts to write powerful, modular, robust and auto-documenting workflows with full provenance automatically stored as they run.
With the integration of AiiDA's plugin system, sharing these workflows with the public is made simple, enabling the wider community to reuse the scientific knowledge encoded within.
While the engine is particularly well suited to manage high-throughput workflows whose steps involve simulations running on high-performance computing infrastructures, it is versatile enough to also effortlessly run code on smaller machines such as personal desktops.

With the release of AiiDA 1.0, its extensible and modular nature has gained adoption within the materials discovery community with over 100 supported simulation code executables and many workflows \footnote{\url{https://aiidateam.github.io/aiida-registry/}} available at the time of writing, many of which have contributed directly to published scientific works.
These are all encouraging signs, as the ultimate goal of AiiDA is to provide the community with a useful tool that can be used as part of an interoperable, FAIR, computational infrastructure to accelerate scientific discovery.
As the scientific community transitions to the exascale era, there is little doubt that such tools will have a greater and greater role to play in the daily activities of researchers.

%% file: sections/acknowledgments.tex
The authors thank Dominik Gresch, Tiziano M\"{u}ller and Jens Br\"{o}der for detailed technical discussions.  We thank the whole AiiDA team for contributions throughout the development process.

MU would like to thank Matthieu Mottet for valuable discussions regarding asyncio and coroutines in Python.

We thanks the Swiss Data Science Center for a stimulating exchange regarding data provenance frameworks.

This work is supported by the MARVEL National Centre for Competency in Research funded by the Swiss National Science Foundation (grant agreement ID 51NF40-182892), the European Centre of Excellence MaX ``Materials design at the Exascale'' (grant number 824143), by the Swiss Platform for Advanced Scientific Computing (PASC) and by the swissuniversities P-5 ``Materials Cloud'' project (grant agreement ID 182-008).
This work was supported by grants from the Swiss National Supercomputing Centre (CSCS) under project ID s836.
We acknowledge PRACE for awarding us access to Piz Daint at CSCS, Switzerland under grant number 2016153543.

%% file: template.bbl
\begin{thebibliography}{39}%
\makeatletter
\providecommand \@ifxundefined [1]{%
 \@ifx{#1\undefined}
}%
\providecommand \@ifnum [1]{%
 \ifnum #1\expandafter \@firstoftwo
 \else \expandafter \@secondoftwo
 \fi
}%
\providecommand \@ifx [1]{%
 \ifx #1\expandafter \@firstoftwo
 \else \expandafter \@secondoftwo
 \fi
}%
\providecommand \natexlab [1]{#1}%
\providecommand \enquote  [1]{``#1''}%
\providecommand \bibnamefont  [1]{#1}%
\providecommand \bibfnamefont [1]{#1}%
\providecommand \citenamefont [1]{#1}%
\providecommand \href@noop [0]{\@secondoftwo}%
\providecommand \href [0]{\begingroup \@sanitize@url \@href}%
\providecommand \@href[1]{\@@startlink{#1}\@@href}%
\providecommand \@@href[1]{\endgroup#1\@@endlink}%
\providecommand \@sanitize@url [0]{\catcode `\\12\catcode `\$12\catcode
  `\&12\catcode `\#12\catcode `\^12\catcode `\_12\catcode `\%12\relax}%
\providecommand \@@startlink[1]{}%
\providecommand \@@endlink[0]{}%
\providecommand \url  [0]{\begingroup\@sanitize@url \@url }%
\providecommand \@url [1]{\endgroup\@href {#1}{\urlprefix }}%
\providecommand \urlprefix  [0]{URL }%
\providecommand \Eprint [0]{\href }%
\providecommand \doibase [0]{http://dx.doi.org/}%
\providecommand \selectlanguage [0]{\@gobble}%
\providecommand \bibinfo  [0]{\@secondoftwo}%
\providecommand \bibfield  [0]{\@secondoftwo}%
\providecommand \translation [1]{[#1]}%
\providecommand \BibitemOpen [0]{}%
\providecommand \bibitemStop [0]{}%
\providecommand \bibitemNoStop [0]{.\EOS\space}%
\providecommand \EOS [0]{\spacefactor3000\relax}%
\providecommand \BibitemShut  [1]{\csname bibitem#1\endcsname}%
\let\auto@bib@innerbib\@empty
%</preamble>
\bibitem [{\citenamefont {Talia}(2013)}]{Talia:2013}%
  \BibitemOpen
  \bibfield  {author} {\bibinfo {author} {\bibfnamefont {D.}~\bibnamefont
  {Talia}},\ }\href {\doibase 10.1155/2013/404525} {\bibfield  {journal}
  {\bibinfo  {journal} {{ISRN} Software Engineering}\ }\textbf {\bibinfo
  {volume} {2013}},\ \bibinfo {pages} {1} (\bibinfo {year} {2013})}\BibitemShut
  {NoStop}%
\bibitem [{\citenamefont {Wilkinson}\ \emph {et~al.}(2016)\citenamefont
  {Wilkinson}, \citenamefont {Dumontier}, \citenamefont {Aalbersberg},
  \citenamefont {Appleton}, \citenamefont {Axton}, \citenamefont {Baak},
  \citenamefont {Blomberg}, \citenamefont {Boiten}, \citenamefont
  {da~Silva~Santos}, \citenamefont {Bourne}, \citenamefont {Bouwman},
  \citenamefont {Brookes}, \citenamefont {Clark}, \citenamefont {Crosas},
  \citenamefont {Dillo}, \citenamefont {Dumon}, \citenamefont {Edmunds},
  \citenamefont {Evelo}, \citenamefont {Finkers}, \citenamefont
  {Gonzalez-Beltran}, \citenamefont {Gray}, \citenamefont {Groth},
  \citenamefont {Goble}, \citenamefont {Grethe}, \citenamefont {Heringa},
  \citenamefont {'t~Hoen}, \citenamefont {Hooft}, \citenamefont {Kuhn},
  \citenamefont {Kok}, \citenamefont {Kok}, \citenamefont {Lusher},
  \citenamefont {Martone}, \citenamefont {Mons}, \citenamefont {Packer},
  \citenamefont {Persson}, \citenamefont {Rocca-Serra}, \citenamefont {Roos},
  \citenamefont {van Schaik}, \citenamefont {Sansone}, \citenamefont
  {Schultes}, \citenamefont {Sengstag}, \citenamefont {Slater}, \citenamefont
  {Strawn}, \citenamefont {Swertz}, \citenamefont {Thompson}, \citenamefont
  {van~der Lei}, \citenamefont {van Mulligen}, \citenamefont {Velterop},
  \citenamefont {Waagmeester}, \citenamefont {Wittenburg}, \citenamefont
  {Wolstencroft}, \citenamefont {Zhao},\ and\ \citenamefont
  {Mons}}]{Wilkinson:2016}%
  \BibitemOpen
  \bibfield  {author} {\bibinfo {author} {\bibfnamefont {M.~D.}\ \bibnamefont
  {Wilkinson}}, \bibinfo {author} {\bibfnamefont {M.}~\bibnamefont
  {Dumontier}}, \bibinfo {author} {\bibfnamefont {I.~J.}\ \bibnamefont
  {Aalbersberg}}, \bibinfo {author} {\bibfnamefont {G.}~\bibnamefont
  {Appleton}}, \bibinfo {author} {\bibfnamefont {M.}~\bibnamefont {Axton}},
  \bibinfo {author} {\bibfnamefont {A.}~\bibnamefont {Baak}}, \bibinfo {author}
  {\bibfnamefont {N.}~\bibnamefont {Blomberg}}, \bibinfo {author}
  {\bibfnamefont {J.-W.}\ \bibnamefont {Boiten}}, \bibinfo {author}
  {\bibfnamefont {L.~B.}\ \bibnamefont {da~Silva~Santos}}, \bibinfo {author}
  {\bibfnamefont {P.~E.}\ \bibnamefont {Bourne}}, \bibinfo {author}
  {\bibfnamefont {J.}~\bibnamefont {Bouwman}}, \bibinfo {author} {\bibfnamefont
  {A.~J.}\ \bibnamefont {Brookes}}, \bibinfo {author} {\bibfnamefont
  {T.}~\bibnamefont {Clark}}, \bibinfo {author} {\bibfnamefont
  {M.}~\bibnamefont {Crosas}}, \bibinfo {author} {\bibfnamefont
  {I.}~\bibnamefont {Dillo}}, \bibinfo {author} {\bibfnamefont
  {O.}~\bibnamefont {Dumon}}, \bibinfo {author} {\bibfnamefont
  {S.}~\bibnamefont {Edmunds}}, \bibinfo {author} {\bibfnamefont {C.~T.}\
  \bibnamefont {Evelo}}, \bibinfo {author} {\bibfnamefont {R.}~\bibnamefont
  {Finkers}}, \bibinfo {author} {\bibfnamefont {A.}~\bibnamefont
  {Gonzalez-Beltran}}, \bibinfo {author} {\bibfnamefont {A.~J.}\ \bibnamefont
  {Gray}}, \bibinfo {author} {\bibfnamefont {P.}~\bibnamefont {Groth}},
  \bibinfo {author} {\bibfnamefont {C.}~\bibnamefont {Goble}}, \bibinfo
  {author} {\bibfnamefont {J.~S.}\ \bibnamefont {Grethe}}, \bibinfo {author}
  {\bibfnamefont {J.}~\bibnamefont {Heringa}}, \bibinfo {author} {\bibfnamefont
  {P.~A.}\ \bibnamefont {'t~Hoen}}, \bibinfo {author} {\bibfnamefont
  {R.}~\bibnamefont {Hooft}}, \bibinfo {author} {\bibfnamefont
  {T.}~\bibnamefont {Kuhn}}, \bibinfo {author} {\bibfnamefont {R.}~\bibnamefont
  {Kok}}, \bibinfo {author} {\bibfnamefont {J.}~\bibnamefont {Kok}}, \bibinfo
  {author} {\bibfnamefont {S.~J.}\ \bibnamefont {Lusher}}, \bibinfo {author}
  {\bibfnamefont {M.~E.}\ \bibnamefont {Martone}}, \bibinfo {author}
  {\bibfnamefont {A.}~\bibnamefont {Mons}}, \bibinfo {author} {\bibfnamefont
  {A.~L.}\ \bibnamefont {Packer}}, \bibinfo {author} {\bibfnamefont
  {B.}~\bibnamefont {Persson}}, \bibinfo {author} {\bibfnamefont
  {P.}~\bibnamefont {Rocca-Serra}}, \bibinfo {author} {\bibfnamefont
  {M.}~\bibnamefont {Roos}}, \bibinfo {author} {\bibfnamefont {R.}~\bibnamefont
  {van Schaik}}, \bibinfo {author} {\bibfnamefont {S.-A.}\ \bibnamefont
  {Sansone}}, \bibinfo {author} {\bibfnamefont {E.}~\bibnamefont {Schultes}},
  \bibinfo {author} {\bibfnamefont {T.}~\bibnamefont {Sengstag}}, \bibinfo
  {author} {\bibfnamefont {T.}~\bibnamefont {Slater}}, \bibinfo {author}
  {\bibfnamefont {G.}~\bibnamefont {Strawn}}, \bibinfo {author} {\bibfnamefont
  {M.~A.}\ \bibnamefont {Swertz}}, \bibinfo {author} {\bibfnamefont
  {M.}~\bibnamefont {Thompson}}, \bibinfo {author} {\bibfnamefont
  {J.}~\bibnamefont {van~der Lei}}, \bibinfo {author} {\bibfnamefont
  {E.}~\bibnamefont {van Mulligen}}, \bibinfo {author} {\bibfnamefont
  {J.}~\bibnamefont {Velterop}}, \bibinfo {author} {\bibfnamefont
  {A.}~\bibnamefont {Waagmeester}}, \bibinfo {author} {\bibfnamefont
  {P.}~\bibnamefont {Wittenburg}}, \bibinfo {author} {\bibfnamefont
  {K.}~\bibnamefont {Wolstencroft}}, \bibinfo {author} {\bibfnamefont
  {J.}~\bibnamefont {Zhao}}, \ and\ \bibinfo {author} {\bibfnamefont
  {B.}~\bibnamefont {Mons}},\ }\href {\doibase 10.1038/sdata.2016.18}
  {\bibfield  {journal} {\bibinfo  {journal} {Scientific Data}\ }\textbf
  {\bibinfo {volume} {3}} (\bibinfo {year} {2016}),\
  10.1038/sdata.2016.18}\BibitemShut {NoStop}%
\bibitem [{\citenamefont {Ioannidis}\ \emph {et~al.}(2009)\citenamefont
  {Ioannidis}, \citenamefont {Allison}, \citenamefont {Ball}, \citenamefont
  {Coulibaly}, \citenamefont {Cui}, \citenamefont {Culhane}, \citenamefont
  {Falchi}, \citenamefont {Furlanello}, \citenamefont {Game}, \citenamefont
  {Jurman}, \citenamefont {Mangion}, \citenamefont {Mehta}, \citenamefont
  {Nitzberg}, \citenamefont {Page}, \citenamefont {Petretto},\ and\
  \citenamefont {van Noort}}]{Ioannidis:2009}%
  \BibitemOpen
  \bibfield  {author} {\bibinfo {author} {\bibfnamefont {J.~P.~A.}\
  \bibnamefont {Ioannidis}}, \bibinfo {author} {\bibfnamefont {D.~B.}\
  \bibnamefont {Allison}}, \bibinfo {author} {\bibfnamefont {C.~A.}\
  \bibnamefont {Ball}}, \bibinfo {author} {\bibfnamefont {I.}~\bibnamefont
  {Coulibaly}}, \bibinfo {author} {\bibfnamefont {X.}~\bibnamefont {Cui}},
  \bibinfo {author} {\bibfnamefont {A.~C.}\ \bibnamefont {Culhane}}, \bibinfo
  {author} {\bibfnamefont {M.}~\bibnamefont {Falchi}}, \bibinfo {author}
  {\bibfnamefont {C.}~\bibnamefont {Furlanello}}, \bibinfo {author}
  {\bibfnamefont {L.}~\bibnamefont {Game}}, \bibinfo {author} {\bibfnamefont
  {G.}~\bibnamefont {Jurman}}, \bibinfo {author} {\bibfnamefont
  {J.}~\bibnamefont {Mangion}}, \bibinfo {author} {\bibfnamefont
  {T.}~\bibnamefont {Mehta}}, \bibinfo {author} {\bibfnamefont
  {M.}~\bibnamefont {Nitzberg}}, \bibinfo {author} {\bibfnamefont {G.~P.}\
  \bibnamefont {Page}}, \bibinfo {author} {\bibfnamefont {E.}~\bibnamefont
  {Petretto}}, \ and\ \bibinfo {author} {\bibfnamefont {V.}~\bibnamefont {van
  Noort}},\ }\href {\doibase 10.1038/ng.295} {\bibfield  {journal} {\bibinfo
  {journal} {Nature Genetics}\ }\textbf {\bibinfo {volume} {41}},\ \bibinfo
  {pages} {149} (\bibinfo {year} {2009})}\BibitemShut {NoStop}%
\bibitem [{\citenamefont {Peng}(2011)}]{Peng:2011}%
  \BibitemOpen
  \bibfield  {author} {\bibinfo {author} {\bibfnamefont {R.~D.}\ \bibnamefont
  {Peng}},\ }\href {\doibase 10.1126/science.1213847} {\bibfield  {journal}
  {\bibinfo  {journal} {Science}\ }\textbf {\bibinfo {volume} {334}},\ \bibinfo
  {pages} {1226} (\bibinfo {year} {2011})}\BibitemShut {NoStop}%
\bibitem [{\citenamefont {Stoddart}(2016)}]{Stoddart:2016}%
  \BibitemOpen
  \bibfield  {author} {\bibinfo {author} {\bibfnamefont {C.}~\bibnamefont
  {Stoddart}},\ }\href {\doibase 10.1038/d41586-019-00067-3} {\bibfield
  {journal} {\bibinfo  {journal} {Nature}\ } (\bibinfo {year} {2016}),\
  10.1038/d41586-019-00067-3}\BibitemShut {NoStop}%
\bibitem [{\citenamefont {Allison}\ \emph {et~al.}(2016)\citenamefont
  {Allison}, \citenamefont {Brown}, \citenamefont {George},\ and\ \citenamefont
  {Kaiser}}]{Allison:2016}%
  \BibitemOpen
  \bibfield  {author} {\bibinfo {author} {\bibfnamefont {D.~B.}\ \bibnamefont
  {Allison}}, \bibinfo {author} {\bibfnamefont {A.~W.}\ \bibnamefont {Brown}},
  \bibinfo {author} {\bibfnamefont {B.~J.}\ \bibnamefont {George}}, \ and\
  \bibinfo {author} {\bibfnamefont {K.~A.}\ \bibnamefont {Kaiser}},\ }\href
  {\doibase 10.1038/530027a} {\bibfield  {journal} {\bibinfo  {journal}
  {Nature}\ }\textbf {\bibinfo {volume} {530}},\ \bibinfo {pages} {27}
  (\bibinfo {year} {2016})}\BibitemShut {NoStop}%
\bibitem [{\citenamefont {Goble}\ \emph {et~al.}(2020)\citenamefont {Goble},
  \citenamefont {Cohen-Boulakia}, \citenamefont {Soiland-Reyes}, \citenamefont
  {Garijo}, \citenamefont {Gil}, \citenamefont {Crusoe}, \citenamefont
  {Peters},\ and\ \citenamefont {Schober}}]{Goble:2020}%
  \BibitemOpen
  \bibfield  {author} {\bibinfo {author} {\bibfnamefont {C.}~\bibnamefont
  {Goble}}, \bibinfo {author} {\bibfnamefont {S.}~\bibnamefont
  {Cohen-Boulakia}}, \bibinfo {author} {\bibfnamefont {S.}~\bibnamefont
  {Soiland-Reyes}}, \bibinfo {author} {\bibfnamefont {D.}~\bibnamefont
  {Garijo}}, \bibinfo {author} {\bibfnamefont {Y.}~\bibnamefont {Gil}},
  \bibinfo {author} {\bibfnamefont {M.~R.}\ \bibnamefont {Crusoe}}, \bibinfo
  {author} {\bibfnamefont {K.}~\bibnamefont {Peters}}, \ and\ \bibinfo {author}
  {\bibfnamefont {D.}~\bibnamefont {Schober}},\ }\href {\doibase
  10.1162/dint_a_00033} {\bibfield  {journal} {\bibinfo  {journal} {Data
  Intelligence}\ }\textbf {\bibinfo {volume} {2}},\ \bibinfo {pages} {108}
  (\bibinfo {year} {2020})}\BibitemShut {NoStop}%
\bibitem [{\citenamefont {Huber}\ \emph {et~al.}(2020)\citenamefont {Huber},
  \citenamefont {Zoupanos}, \citenamefont {Uhrin}, \citenamefont {Talirz},
  \citenamefont {Kahle}, \citenamefont {H\"auselmann}, \citenamefont {Gresch},
  \citenamefont {M\"uller}, \citenamefont {Yakutovich}, \citenamefont
  {Andersen}, \citenamefont {Ramirez}, \citenamefont {Adorf}, \citenamefont
  {Gargiulo}, \citenamefont {Kumbhar}, \citenamefont {Passaro}, \citenamefont
  {Johnston}, \citenamefont {Merkys}, \citenamefont {Cepellotti}, \citenamefont
  {Mounet}, \citenamefont {Marzari}, \citenamefont {Kozinsky},\ and\
  \citenamefont {Pizzi}}]{Huber:2020}%
  \BibitemOpen
  \bibfield  {author} {\bibinfo {author} {\bibfnamefont {S.~P.}\ \bibnamefont
  {Huber}}, \bibinfo {author} {\bibfnamefont {S.}~\bibnamefont {Zoupanos}},
  \bibinfo {author} {\bibfnamefont {M.}~\bibnamefont {Uhrin}}, \bibinfo
  {author} {\bibfnamefont {L.}~\bibnamefont {Talirz}}, \bibinfo {author}
  {\bibfnamefont {L.}~\bibnamefont {Kahle}}, \bibinfo {author} {\bibfnamefont
  {R.}~\bibnamefont {H\"auselmann}}, \bibinfo {author} {\bibfnamefont
  {D.}~\bibnamefont {Gresch}}, \bibinfo {author} {\bibfnamefont
  {T.}~\bibnamefont {M\"uller}}, \bibinfo {author} {\bibfnamefont {A.~V.}\
  \bibnamefont {Yakutovich}}, \bibinfo {author} {\bibfnamefont {C.~W.}\
  \bibnamefont {Andersen}}, \bibinfo {author} {\bibfnamefont {F.~F.}\
  \bibnamefont {Ramirez}}, \bibinfo {author} {\bibfnamefont {C.~S.}\
  \bibnamefont {Adorf}}, \bibinfo {author} {\bibfnamefont {F.}~\bibnamefont
  {Gargiulo}}, \bibinfo {author} {\bibfnamefont {S.}~\bibnamefont {Kumbhar}},
  \bibinfo {author} {\bibfnamefont {E.}~\bibnamefont {Passaro}}, \bibinfo
  {author} {\bibfnamefont {C.}~\bibnamefont {Johnston}}, \bibinfo {author}
  {\bibfnamefont {A.}~\bibnamefont {Merkys}}, \bibinfo {author} {\bibfnamefont
  {A.}~\bibnamefont {Cepellotti}}, \bibinfo {author} {\bibfnamefont
  {N.}~\bibnamefont {Mounet}}, \bibinfo {author} {\bibfnamefont
  {N.}~\bibnamefont {Marzari}}, \bibinfo {author} {\bibfnamefont
  {B.}~\bibnamefont {Kozinsky}}, \ and\ \bibinfo {author} {\bibfnamefont
  {G.}~\bibnamefont {Pizzi}},\ }\href {https://arxiv.org/abs/1808.01590}
  {\bibfield  {journal} {\bibinfo  {journal} {ArXiv e-prints}\ } (\bibinfo
  {year} {2020})},\ \Eprint {http://arxiv.org/abs/2003.12476} {arXiv:2003.12476
  [cs.DC]} \BibitemShut {NoStop}%
\bibitem [{\citenamefont {Altintas}\ \emph {et~al.}()\citenamefont {Altintas},
  \citenamefont {Berkley}, \citenamefont {Jaeger}, \citenamefont {Jones},
  \citenamefont {Ludascher},\ and\ \citenamefont {Mock}}]{Altintas}%
  \BibitemOpen
  \bibfield  {author} {\bibinfo {author} {\bibfnamefont {I.}~\bibnamefont
  {Altintas}}, \bibinfo {author} {\bibfnamefont {C.}~\bibnamefont {Berkley}},
  \bibinfo {author} {\bibfnamefont {E.}~\bibnamefont {Jaeger}}, \bibinfo
  {author} {\bibfnamefont {M.}~\bibnamefont {Jones}}, \bibinfo {author}
  {\bibfnamefont {B.}~\bibnamefont {Ludascher}}, \ and\ \bibinfo {author}
  {\bibfnamefont {S.}~\bibnamefont {Mock}},\ }in\ \href {\doibase
  10.1109/ssdm.2004.1311241} {\emph {\bibinfo {booktitle} {Proceedings. 16th
  International Conference on Scientific and Statistical Database Management,
  2004.}}}\ (\bibinfo  {publisher} {{IEEE}})\BibitemShut {NoStop}%
\bibitem [{\citenamefont {Oinn}\ \emph {et~al.}(2004)\citenamefont {Oinn},
  \citenamefont {Addis}, \citenamefont {Ferris}, \citenamefont {Marvin},
  \citenamefont {Senger}, \citenamefont {Greenwood}, \citenamefont {Carver},
  \citenamefont {Glover}, \citenamefont {Pocock}, \citenamefont {Wipat},\ and\
  \citenamefont {Li}}]{Oinn:2004}%
  \BibitemOpen
  \bibfield  {author} {\bibinfo {author} {\bibfnamefont {T.}~\bibnamefont
  {Oinn}}, \bibinfo {author} {\bibfnamefont {M.}~\bibnamefont {Addis}},
  \bibinfo {author} {\bibfnamefont {J.}~\bibnamefont {Ferris}}, \bibinfo
  {author} {\bibfnamefont {D.}~\bibnamefont {Marvin}}, \bibinfo {author}
  {\bibfnamefont {M.}~\bibnamefont {Senger}}, \bibinfo {author} {\bibfnamefont
  {M.}~\bibnamefont {Greenwood}}, \bibinfo {author} {\bibfnamefont
  {T.}~\bibnamefont {Carver}}, \bibinfo {author} {\bibfnamefont
  {K.}~\bibnamefont {Glover}}, \bibinfo {author} {\bibfnamefont {M.~R.}\
  \bibnamefont {Pocock}}, \bibinfo {author} {\bibfnamefont {A.}~\bibnamefont
  {Wipat}}, \ and\ \bibinfo {author} {\bibfnamefont {P.}~\bibnamefont {Li}},\
  }\href {\doibase 10.1093/bioinformatics/bth361} {\bibfield  {journal}
  {\bibinfo  {journal} {Bioinformatics}\ }\textbf {\bibinfo {volume} {20}},\
  \bibinfo {pages} {3045} (\bibinfo {year} {2004})}\BibitemShut {NoStop}%
\bibitem [{\citenamefont {Taylor}\ \emph {et~al.}(2003)\citenamefont {Taylor},
  \citenamefont {Shields}, \citenamefont {Wang},\ and\ \citenamefont
  {Rana}}]{Taylor:2003}%
  \BibitemOpen
  \bibfield  {author} {\bibinfo {author} {\bibfnamefont {I.}~\bibnamefont
  {Taylor}}, \bibinfo {author} {\bibfnamefont {M.}~\bibnamefont {Shields}},
  \bibinfo {author} {\bibfnamefont {I.}~\bibnamefont {Wang}}, \ and\ \bibinfo
  {author} {\bibfnamefont {O.}~\bibnamefont {Rana}},\ }\href {\doibase
  10.1023/b:grid.0000024074.63139.ce} {\bibfield  {journal} {\bibinfo
  {journal} {Journal of Grid Computing}\ }\textbf {\bibinfo {volume} {1}},\
  \bibinfo {pages} {199} (\bibinfo {year} {2003})}\BibitemShut {NoStop}%
\bibitem [{\citenamefont {von Laszewski}\ \emph {et~al.}(2007)\citenamefont
  {von Laszewski}, \citenamefont {Hategan},\ and\ \citenamefont
  {Kodeboyina}}]{vonLaszewski:2007}%
  \BibitemOpen
  \bibfield  {author} {\bibinfo {author} {\bibfnamefont {G.}~\bibnamefont {von
  Laszewski}}, \bibinfo {author} {\bibfnamefont {M.}~\bibnamefont {Hategan}}, \
  and\ \bibinfo {author} {\bibfnamefont {D.}~\bibnamefont {Kodeboyina}},\ }in\
  \href {\doibase 10.1007/978-1-84628-757-2_21} {\emph {\bibinfo {booktitle}
  {Workflows for e-Science}}}\ (\bibinfo  {publisher} {Springer London},\
  \bibinfo {year} {2007})\ pp.\ \bibinfo {pages} {340--356}\BibitemShut
  {NoStop}%
\bibitem [{\citenamefont {Fahringer}\ \emph
  {et~al.}(2005{\natexlab{a}})\citenamefont {Fahringer}, \citenamefont
  {Prodan}, \citenamefont {Duan}, \citenamefont {Nerieri}, \citenamefont
  {Podlipnig}, \citenamefont {Qin}, \citenamefont {Siddiqui}, \citenamefont
  {Truong}, \citenamefont {Villazon},\ and\ \citenamefont
  {Wieczorek}}]{Fahringer:2005}%
  \BibitemOpen
  \bibfield  {author} {\bibinfo {author} {\bibfnamefont {T.}~\bibnamefont
  {Fahringer}}, \bibinfo {author} {\bibfnamefont {R.}~\bibnamefont {Prodan}},
  \bibinfo {author} {\bibfnamefont {R.}~\bibnamefont {Duan}}, \bibinfo {author}
  {\bibfnamefont {F.}~\bibnamefont {Nerieri}}, \bibinfo {author} {\bibfnamefont
  {S.}~\bibnamefont {Podlipnig}}, \bibinfo {author} {\bibfnamefont
  {J.}~\bibnamefont {Qin}}, \bibinfo {author} {\bibfnamefont {M.}~\bibnamefont
  {Siddiqui}}, \bibinfo {author} {\bibfnamefont {H.-L.}\ \bibnamefont
  {Truong}}, \bibinfo {author} {\bibfnamefont {A.}~\bibnamefont {Villazon}}, \
  and\ \bibinfo {author} {\bibfnamefont {M.}~\bibnamefont {Wieczorek}},\ }in\
  \href {\doibase 10.1109/grid.2005.1542733} {\emph {\bibinfo {booktitle} {The
  6th {IEEE}/{ACM} International Workshop on Grid Computing, 2005.}}}\
  (\bibinfo  {publisher} {{IEEE}},\ \bibinfo {year} {2005})\BibitemShut
  {NoStop}%
\bibitem [{\citenamefont {Fahringer}\ \emph
  {et~al.}(2005{\natexlab{b}})\citenamefont {Fahringer}, \citenamefont {Qin},\
  and\ \citenamefont {Hainzer}}]{FahringerHainzer:2005}%
  \BibitemOpen
  \bibfield  {author} {\bibinfo {author} {\bibfnamefont {T.}~\bibnamefont
  {Fahringer}}, \bibinfo {author} {\bibfnamefont {J.}~\bibnamefont {Qin}}, \
  and\ \bibinfo {author} {\bibfnamefont {S.}~\bibnamefont {Hainzer}},\ }in\
  \href {\doibase 10.1109/ccgrid.2005.1558629} {\emph {\bibinfo {booktitle}
  {{CCGrid} 2005. {IEEE} International Symposium on Cluster Computing and the
  Grid, 2005.}}}\ (\bibinfo  {publisher} {{IEEE}},\ \bibinfo {year}
  {2005})\BibitemShut {NoStop}%
\bibitem [{\citenamefont {Chapman}\ \emph {et~al.}(2016)\citenamefont
  {Chapman}, \citenamefont {Chilton}, \citenamefont {Heuer}, \citenamefont
  {Kartashov}, \citenamefont {Leehr}, \citenamefont {M{\'e}nager},
  \citenamefont {Nedeljkovich}, \citenamefont {Scales}, \citenamefont
  {Soiland-Reyes},\ and\ \citenamefont {Stojanovic}}]{Amstutz:2016}%
  \BibitemOpen
  \bibfield  {author} {\bibinfo {author} {\bibfnamefont {B.}~\bibnamefont
  {Chapman}}, \bibinfo {author} {\bibfnamefont {J.}~\bibnamefont {Chilton}},
  \bibinfo {author} {\bibfnamefont {M.}~\bibnamefont {Heuer}}, \bibinfo
  {author} {\bibfnamefont {A.}~\bibnamefont {Kartashov}}, \bibinfo {author}
  {\bibfnamefont {D.}~\bibnamefont {Leehr}}, \bibinfo {author} {\bibfnamefont
  {H.}~\bibnamefont {M{\'e}nager}}, \bibinfo {author} {\bibfnamefont
  {M.}~\bibnamefont {Nedeljkovich}}, \bibinfo {author} {\bibfnamefont
  {M.}~\bibnamefont {Scales}}, \bibinfo {author} {\bibfnamefont
  {S.}~\bibnamefont {Soiland-Reyes}}, \ and\ \bibinfo {author} {\bibfnamefont
  {L.}~\bibnamefont {Stojanovic}},\ }\href {\doibase
  10.6084/m9.figshare.3115156.v2} {\emph {\bibinfo {title} {Common Workflow
  Language, v1.0}}}\ (\bibinfo  {publisher} {figshare},\ \bibinfo {year}
  {2016})\BibitemShut {NoStop}%
\bibitem [{\citenamefont {Deelman}\ \emph {et~al.}(2005)\citenamefont
  {Deelman}, \citenamefont {Singh}, \citenamefont {Su}, \citenamefont {Blythe},
  \citenamefont {Gil}, \citenamefont {Kesselman}, \citenamefont {Mehta},
  \citenamefont {Vahi}, \citenamefont {Berriman}, \citenamefont {Good},
  \citenamefont {Laity}, \citenamefont {Jacob},\ and\ \citenamefont
  {Katz}}]{Deelman:2005}%
  \BibitemOpen
  \bibfield  {author} {\bibinfo {author} {\bibfnamefont {E.}~\bibnamefont
  {Deelman}}, \bibinfo {author} {\bibfnamefont {G.}~\bibnamefont {Singh}},
  \bibinfo {author} {\bibfnamefont {M.-H.}\ \bibnamefont {Su}}, \bibinfo
  {author} {\bibfnamefont {J.}~\bibnamefont {Blythe}}, \bibinfo {author}
  {\bibfnamefont {Y.}~\bibnamefont {Gil}}, \bibinfo {author} {\bibfnamefont
  {C.}~\bibnamefont {Kesselman}}, \bibinfo {author} {\bibfnamefont
  {G.}~\bibnamefont {Mehta}}, \bibinfo {author} {\bibfnamefont
  {K.}~\bibnamefont {Vahi}}, \bibinfo {author} {\bibfnamefont {G.~B.}\
  \bibnamefont {Berriman}}, \bibinfo {author} {\bibfnamefont {J.}~\bibnamefont
  {Good}}, \bibinfo {author} {\bibfnamefont {A.}~\bibnamefont {Laity}},
  \bibinfo {author} {\bibfnamefont {J.~C.}\ \bibnamefont {Jacob}}, \ and\
  \bibinfo {author} {\bibfnamefont {D.~S.}\ \bibnamefont {Katz}},\ }\href
  {\doibase 10.1155/2005/128026} {\bibfield  {journal} {\bibinfo  {journal}
  {Scientific Programming}\ }\textbf {\bibinfo {volume} {13}},\ \bibinfo
  {pages} {219} (\bibinfo {year} {2005})}\BibitemShut {NoStop}%
\bibitem [{Note1()}]{Note1}%
  \BibitemOpen
  \bibinfo {note} {Logic and loops can often be used but these need to be
  represented explicitly which quickly becomes cumbersome.}\BibitemShut {Stop}%
\bibitem [{\citenamefont {Adorf}\ \emph {et~al.}(2018)\citenamefont {Adorf},
  \citenamefont {Dodd}, \citenamefont {Ramasubramani},\ and\ \citenamefont
  {Glotzer}}]{Adorf:2018}%
  \BibitemOpen
  \bibfield  {author} {\bibinfo {author} {\bibfnamefont {C.~S.}\ \bibnamefont
  {Adorf}}, \bibinfo {author} {\bibfnamefont {P.~M.}\ \bibnamefont {Dodd}},
  \bibinfo {author} {\bibfnamefont {V.}~\bibnamefont {Ramasubramani}}, \ and\
  \bibinfo {author} {\bibfnamefont {S.~C.}\ \bibnamefont {Glotzer}},\ }\href
  {\doibase 10.1016/j.commatsci.2018.01.035} {\bibfield  {journal} {\bibinfo
  {journal} {Computational Materials Science}\ }\textbf {\bibinfo {volume}
  {146}},\ \bibinfo {pages} {220} (\bibinfo {year} {2018})}\BibitemShut
  {NoStop}%
\bibitem [{\citenamefont {Babuji}\ \emph {et~al.}(2019)\citenamefont {Babuji},
  \citenamefont {Foster}, \citenamefont {Wilde}, \citenamefont {Chard},
  \citenamefont {Woodard}, \citenamefont {Li}, \citenamefont {Katz},
  \citenamefont {Clifford}, \citenamefont {Kumar}, \citenamefont {Lacinski},
  \citenamefont {Chard},\ and\ \citenamefont {Wozniak}}]{Babuji:2019}%
  \BibitemOpen
  \bibfield  {author} {\bibinfo {author} {\bibfnamefont {Y.}~\bibnamefont
  {Babuji}}, \bibinfo {author} {\bibfnamefont {I.}~\bibnamefont {Foster}},
  \bibinfo {author} {\bibfnamefont {M.}~\bibnamefont {Wilde}}, \bibinfo
  {author} {\bibfnamefont {K.}~\bibnamefont {Chard}}, \bibinfo {author}
  {\bibfnamefont {A.}~\bibnamefont {Woodard}}, \bibinfo {author} {\bibfnamefont
  {Z.}~\bibnamefont {Li}}, \bibinfo {author} {\bibfnamefont {D.~S.}\
  \bibnamefont {Katz}}, \bibinfo {author} {\bibfnamefont {B.}~\bibnamefont
  {Clifford}}, \bibinfo {author} {\bibfnamefont {R.}~\bibnamefont {Kumar}},
  \bibinfo {author} {\bibfnamefont {L.}~\bibnamefont {Lacinski}}, \bibinfo
  {author} {\bibfnamefont {R.}~\bibnamefont {Chard}}, \ and\ \bibinfo {author}
  {\bibfnamefont {J.~M.}\ \bibnamefont {Wozniak}},\ }in\ \href {\doibase
  10.1145/3307681.3325400} {\emph {\bibinfo {booktitle} {Proceedings of the
  28th International Symposium on High-Performance Parallel and Distributed
  Computing - {HPDC} 2019}}}\ (\bibinfo  {publisher} {{ACM} Press},\ \bibinfo
  {year} {2019})\BibitemShut {NoStop}%
\bibitem [{\citenamefont {Jain}\ \emph {et~al.}(2015)\citenamefont {Jain},
  \citenamefont {Ong}, \citenamefont {Chen}, \citenamefont {Medasani},
  \citenamefont {Qu}, \citenamefont {Kocher}, \citenamefont {Brafman},
  \citenamefont {Petretto}, \citenamefont {Rignanese}, \citenamefont {Hautier},
  \citenamefont {Gunter},\ and\ \citenamefont {Persson}}]{Jain:2015}%
  \BibitemOpen
  \bibfield  {author} {\bibinfo {author} {\bibfnamefont {A.}~\bibnamefont
  {Jain}}, \bibinfo {author} {\bibfnamefont {S.~P.}\ \bibnamefont {Ong}},
  \bibinfo {author} {\bibfnamefont {W.}~\bibnamefont {Chen}}, \bibinfo {author}
  {\bibfnamefont {B.}~\bibnamefont {Medasani}}, \bibinfo {author}
  {\bibfnamefont {X.}~\bibnamefont {Qu}}, \bibinfo {author} {\bibfnamefont
  {M.}~\bibnamefont {Kocher}}, \bibinfo {author} {\bibfnamefont
  {M.}~\bibnamefont {Brafman}}, \bibinfo {author} {\bibfnamefont
  {G.}~\bibnamefont {Petretto}}, \bibinfo {author} {\bibfnamefont {G.-M.}\
  \bibnamefont {Rignanese}}, \bibinfo {author} {\bibfnamefont {G.}~\bibnamefont
  {Hautier}}, \bibinfo {author} {\bibfnamefont {D.}~\bibnamefont {Gunter}}, \
  and\ \bibinfo {author} {\bibfnamefont {K.~A.}\ \bibnamefont {Persson}},\
  }\href {\doibase 10.1002/cpe.3505} {\bibfield  {journal} {\bibinfo  {journal}
  {Concurrency and Computation: Practice and Experience}\ }\textbf {\bibinfo
  {volume} {27}},\ \bibinfo {pages} {5037} (\bibinfo {year}
  {2015})}\BibitemShut {NoStop}%
\bibitem [{Bra(2014)}]{Bray:2014}%
  \BibitemOpen
  \href {\doibase 10.17487/rfc7159} {\emph {\bibinfo {title} {The {JavaScript}
  Object Notation ({JSON}) Data Interchange Format}}},\ \bibinfo {type} {Tech.
  Rep.}\ (\bibinfo {year} {2014})\BibitemShut {NoStop}%
\bibitem [{\citenamefont {Curtarolo}\ \emph {et~al.}(2012)\citenamefont
  {Curtarolo}, \citenamefont {Setyawan}, \citenamefont {Hart}, \citenamefont
  {Jahnatek}, \citenamefont {Chepulskii}, \citenamefont {Taylor}, \citenamefont
  {Wang}, \citenamefont {Xue}, \citenamefont {Yang}, \citenamefont {Levy},
  \citenamefont {Mehl}, \citenamefont {Stokes}, \citenamefont {Demchenko},\
  and\ \citenamefont {Morgan}}]{Curtarolo:2012}%
  \BibitemOpen
  \bibfield  {author} {\bibinfo {author} {\bibfnamefont {S.}~\bibnamefont
  {Curtarolo}}, \bibinfo {author} {\bibfnamefont {W.}~\bibnamefont {Setyawan}},
  \bibinfo {author} {\bibfnamefont {G.~L.}\ \bibnamefont {Hart}}, \bibinfo
  {author} {\bibfnamefont {M.}~\bibnamefont {Jahnatek}}, \bibinfo {author}
  {\bibfnamefont {R.~V.}\ \bibnamefont {Chepulskii}}, \bibinfo {author}
  {\bibfnamefont {R.~H.}\ \bibnamefont {Taylor}}, \bibinfo {author}
  {\bibfnamefont {S.}~\bibnamefont {Wang}}, \bibinfo {author} {\bibfnamefont
  {J.}~\bibnamefont {Xue}}, \bibinfo {author} {\bibfnamefont {K.}~\bibnamefont
  {Yang}}, \bibinfo {author} {\bibfnamefont {O.}~\bibnamefont {Levy}}, \bibinfo
  {author} {\bibfnamefont {M.~J.}\ \bibnamefont {Mehl}}, \bibinfo {author}
  {\bibfnamefont {H.~T.}\ \bibnamefont {Stokes}}, \bibinfo {author}
  {\bibfnamefont {D.~O.}\ \bibnamefont {Demchenko}}, \ and\ \bibinfo {author}
  {\bibfnamefont {D.}~\bibnamefont {Morgan}},\ }\href {\doibase
  10.1016/j.commatsci.2012.02.005} {\bibfield  {journal} {\bibinfo  {journal}
  {Computational Materials Science}\ }\textbf {\bibinfo {volume} {58}},\
  \bibinfo {pages} {218} (\bibinfo {year} {2012})}\BibitemShut {NoStop}%
\bibitem [{\citenamefont {Mathew}\ \emph {et~al.}(2017)\citenamefont {Mathew},
  \citenamefont {Montoya}, \citenamefont {Faghaninia}, \citenamefont
  {Dwarakanath}, \citenamefont {Aykol}, \citenamefont {Tang}, \citenamefont
  {heng Chu}, \citenamefont {Smidt}, \citenamefont {Bocklund}, \citenamefont
  {Horton}, \citenamefont {Dagdelen}, \citenamefont {Wood}, \citenamefont
  {Liu}, \citenamefont {Neaton}, \citenamefont {Ong}, \citenamefont {Persson},\
  and\ \citenamefont {Jain}}]{Mathew:2017}%
  \BibitemOpen
  \bibfield  {author} {\bibinfo {author} {\bibfnamefont {K.}~\bibnamefont
  {Mathew}}, \bibinfo {author} {\bibfnamefont {J.~H.}\ \bibnamefont {Montoya}},
  \bibinfo {author} {\bibfnamefont {A.}~\bibnamefont {Faghaninia}}, \bibinfo
  {author} {\bibfnamefont {S.}~\bibnamefont {Dwarakanath}}, \bibinfo {author}
  {\bibfnamefont {M.}~\bibnamefont {Aykol}}, \bibinfo {author} {\bibfnamefont
  {H.}~\bibnamefont {Tang}}, \bibinfo {author} {\bibfnamefont {I.}~\bibnamefont
  {heng Chu}}, \bibinfo {author} {\bibfnamefont {T.}~\bibnamefont {Smidt}},
  \bibinfo {author} {\bibfnamefont {B.}~\bibnamefont {Bocklund}}, \bibinfo
  {author} {\bibfnamefont {M.}~\bibnamefont {Horton}}, \bibinfo {author}
  {\bibfnamefont {J.}~\bibnamefont {Dagdelen}}, \bibinfo {author}
  {\bibfnamefont {B.}~\bibnamefont {Wood}}, \bibinfo {author} {\bibfnamefont
  {Z.-K.}\ \bibnamefont {Liu}}, \bibinfo {author} {\bibfnamefont
  {J.}~\bibnamefont {Neaton}}, \bibinfo {author} {\bibfnamefont {S.~P.}\
  \bibnamefont {Ong}}, \bibinfo {author} {\bibfnamefont {K.}~\bibnamefont
  {Persson}}, \ and\ \bibinfo {author} {\bibfnamefont {A.}~\bibnamefont
  {Jain}},\ }\href {\doibase 10.1016/j.commatsci.2017.07.030} {\bibfield
  {journal} {\bibinfo  {journal} {Computational Materials Science}\ }\textbf
  {\bibinfo {volume} {139}},\ \bibinfo {pages} {140} (\bibinfo {year}
  {2017})}\BibitemShut {NoStop}%
\bibitem [{\citenamefont {Mayeshiba}\ \emph {et~al.}(2017)\citenamefont
  {Mayeshiba}, \citenamefont {Wu}, \citenamefont {Angsten}, \citenamefont
  {Kaczmarowski}, \citenamefont {Song}, \citenamefont {Jenness}, \citenamefont
  {Xie},\ and\ \citenamefont {Morgan}}]{Mayeshiba:2017}%
  \BibitemOpen
  \bibfield  {author} {\bibinfo {author} {\bibfnamefont {T.}~\bibnamefont
  {Mayeshiba}}, \bibinfo {author} {\bibfnamefont {H.}~\bibnamefont {Wu}},
  \bibinfo {author} {\bibfnamefont {T.}~\bibnamefont {Angsten}}, \bibinfo
  {author} {\bibfnamefont {A.}~\bibnamefont {Kaczmarowski}}, \bibinfo {author}
  {\bibfnamefont {Z.}~\bibnamefont {Song}}, \bibinfo {author} {\bibfnamefont
  {G.}~\bibnamefont {Jenness}}, \bibinfo {author} {\bibfnamefont
  {W.}~\bibnamefont {Xie}}, \ and\ \bibinfo {author} {\bibfnamefont
  {D.}~\bibnamefont {Morgan}},\ }\href {\doibase
  10.1016/j.commatsci.2016.09.018} {\bibfield  {journal} {\bibinfo  {journal}
  {Computational Materials Science}\ }\textbf {\bibinfo {volume} {126}},\
  \bibinfo {pages} {90} (\bibinfo {year} {2017})}\BibitemShut {NoStop}%
\bibitem [{\citenamefont {Saal}\ \emph {et~al.}(2013)\citenamefont {Saal},
  \citenamefont {Kirklin}, \citenamefont {Aykol}, \citenamefont {Meredig},\
  and\ \citenamefont {Wolverton}}]{Saal:2013}%
  \BibitemOpen
  \bibfield  {author} {\bibinfo {author} {\bibfnamefont {J.~E.}\ \bibnamefont
  {Saal}}, \bibinfo {author} {\bibfnamefont {S.}~\bibnamefont {Kirklin}},
  \bibinfo {author} {\bibfnamefont {M.}~\bibnamefont {Aykol}}, \bibinfo
  {author} {\bibfnamefont {B.}~\bibnamefont {Meredig}}, \ and\ \bibinfo
  {author} {\bibfnamefont {C.}~\bibnamefont {Wolverton}},\ }\href {\doibase
  10.1007/s11837-013-0755-4} {\bibfield  {journal} {\bibinfo  {journal}
  {{JOM}}\ }\textbf {\bibinfo {volume} {65}},\ \bibinfo {pages} {1501}
  (\bibinfo {year} {2013})}\BibitemShut {NoStop}%
\bibitem [{\citenamefont {Lejaeghere}\ \emph {et~al.}(2016)\citenamefont
  {Lejaeghere}, \citenamefont {Bihlmayer}, \citenamefont {Bjorkman},
  \citenamefont {Blaha}, \citenamefont {Blugel}, \citenamefont {Blum},
  \citenamefont {Caliste}, \citenamefont {Castelli}, \citenamefont {Clark},
  \citenamefont {Corso}, \citenamefont {de~Gironcoli}, \citenamefont {Deutsch},
  \citenamefont {Dewhurst}, \citenamefont {Marco}, \citenamefont {Draxl},
  \citenamefont {ak}, \citenamefont {Eriksson}, \citenamefont {Flores-Livas},
  \citenamefont {Garrity}, \citenamefont {Genovese}, \citenamefont {Giannozzi},
  \citenamefont {Giantomassi}, \citenamefont {Goedecker}, \citenamefont
  {Gonze}, \citenamefont {Granas}, \citenamefont {Gross}, \citenamefont
  {Gulans}, \citenamefont {Gygi}, \citenamefont {Hamann}, \citenamefont
  {Hasnip}, \citenamefont {Holzwarth}, \citenamefont {an}, \citenamefont
  {Jochym}, \citenamefont {Jollet}, \citenamefont {Jones}, \citenamefont
  {Kresse}, \citenamefont {Koepernik}, \citenamefont {Kucukbenli},
  \citenamefont {Kvashnin}, \citenamefont {Locht}, \citenamefont {Lubeck},
  \citenamefont {Marsman}, \citenamefont {Marzari}, \citenamefont {Nitzsche},
  \citenamefont {Nordstrom}, \citenamefont {Ozaki}, \citenamefont {Paulatto},
  \citenamefont {Pickard}, \citenamefont {Poelmans}, \citenamefont {Probert},
  \citenamefont {Refson}, \citenamefont {Richter}, \citenamefont {Rignanese},
  \citenamefont {Saha}, \citenamefont {Scheffler}, \citenamefont {Schlipf},
  \citenamefont {Schwarz}, \citenamefont {Sharma}, \citenamefont {Tavazza},
  \citenamefont {Thunstrom}, \citenamefont {Tkatchenko}, \citenamefont
  {Torrent}, \citenamefont {Vanderbilt}, \citenamefont {van Setten},
  \citenamefont {Speybroeck}, \citenamefont {Wills}, \citenamefont {Yates},
  \citenamefont {Zhang},\ and\ \citenamefont {Cottenier}}]{Lejaeghere:2016}%
  \BibitemOpen
  \bibfield  {author} {\bibinfo {author} {\bibfnamefont {K.}~\bibnamefont
  {Lejaeghere}}, \bibinfo {author} {\bibfnamefont {G.}~\bibnamefont
  {Bihlmayer}}, \bibinfo {author} {\bibfnamefont {T.}~\bibnamefont {Bjorkman}},
  \bibinfo {author} {\bibfnamefont {P.}~\bibnamefont {Blaha}}, \bibinfo
  {author} {\bibfnamefont {S.}~\bibnamefont {Blugel}}, \bibinfo {author}
  {\bibfnamefont {V.}~\bibnamefont {Blum}}, \bibinfo {author} {\bibfnamefont
  {D.}~\bibnamefont {Caliste}}, \bibinfo {author} {\bibfnamefont {I.~E.}\
  \bibnamefont {Castelli}}, \bibinfo {author} {\bibfnamefont {S.~J.}\
  \bibnamefont {Clark}}, \bibinfo {author} {\bibfnamefont {A.~D.}\ \bibnamefont
  {Corso}}, \bibinfo {author} {\bibfnamefont {S.}~\bibnamefont {de~Gironcoli}},
  \bibinfo {author} {\bibfnamefont {T.}~\bibnamefont {Deutsch}}, \bibinfo
  {author} {\bibfnamefont {J.~K.}\ \bibnamefont {Dewhurst}}, \bibinfo {author}
  {\bibfnamefont {I.~D.}\ \bibnamefont {Marco}}, \bibinfo {author}
  {\bibfnamefont {C.}~\bibnamefont {Draxl}}, \bibinfo {author} {\bibfnamefont
  {M.~D.}\ \bibnamefont {ak}}, \bibinfo {author} {\bibfnamefont
  {O.}~\bibnamefont {Eriksson}}, \bibinfo {author} {\bibfnamefont {J.~A.}\
  \bibnamefont {Flores-Livas}}, \bibinfo {author} {\bibfnamefont {K.~F.}\
  \bibnamefont {Garrity}}, \bibinfo {author} {\bibfnamefont {L.}~\bibnamefont
  {Genovese}}, \bibinfo {author} {\bibfnamefont {P.}~\bibnamefont {Giannozzi}},
  \bibinfo {author} {\bibfnamefont {M.}~\bibnamefont {Giantomassi}}, \bibinfo
  {author} {\bibfnamefont {S.}~\bibnamefont {Goedecker}}, \bibinfo {author}
  {\bibfnamefont {X.}~\bibnamefont {Gonze}}, \bibinfo {author} {\bibfnamefont
  {O.}~\bibnamefont {Granas}}, \bibinfo {author} {\bibfnamefont {E.~K.~U.}\
  \bibnamefont {Gross}}, \bibinfo {author} {\bibfnamefont {A.}~\bibnamefont
  {Gulans}}, \bibinfo {author} {\bibfnamefont {F.}~\bibnamefont {Gygi}},
  \bibinfo {author} {\bibfnamefont {D.~R.}\ \bibnamefont {Hamann}}, \bibinfo
  {author} {\bibfnamefont {P.~J.}\ \bibnamefont {Hasnip}}, \bibinfo {author}
  {\bibfnamefont {N.~A.~W.}\ \bibnamefont {Holzwarth}}, \bibinfo {author}
  {\bibfnamefont {D.~I.}\ \bibnamefont {an}}, \bibinfo {author} {\bibfnamefont
  {D.~B.}\ \bibnamefont {Jochym}}, \bibinfo {author} {\bibfnamefont
  {F.}~\bibnamefont {Jollet}}, \bibinfo {author} {\bibfnamefont
  {D.}~\bibnamefont {Jones}}, \bibinfo {author} {\bibfnamefont
  {G.}~\bibnamefont {Kresse}}, \bibinfo {author} {\bibfnamefont
  {K.}~\bibnamefont {Koepernik}}, \bibinfo {author} {\bibfnamefont
  {E.}~\bibnamefont {Kucukbenli}}, \bibinfo {author} {\bibfnamefont {Y.~O.}\
  \bibnamefont {Kvashnin}}, \bibinfo {author} {\bibfnamefont {I.~L.~M.}\
  \bibnamefont {Locht}}, \bibinfo {author} {\bibfnamefont {S.}~\bibnamefont
  {Lubeck}}, \bibinfo {author} {\bibfnamefont {M.}~\bibnamefont {Marsman}},
  \bibinfo {author} {\bibfnamefont {N.}~\bibnamefont {Marzari}}, \bibinfo
  {author} {\bibfnamefont {U.}~\bibnamefont {Nitzsche}}, \bibinfo {author}
  {\bibfnamefont {L.}~\bibnamefont {Nordstrom}}, \bibinfo {author}
  {\bibfnamefont {T.}~\bibnamefont {Ozaki}}, \bibinfo {author} {\bibfnamefont
  {L.}~\bibnamefont {Paulatto}}, \bibinfo {author} {\bibfnamefont {C.~J.}\
  \bibnamefont {Pickard}}, \bibinfo {author} {\bibfnamefont {W.}~\bibnamefont
  {Poelmans}}, \bibinfo {author} {\bibfnamefont {M.~I.~J.}\ \bibnamefont
  {Probert}}, \bibinfo {author} {\bibfnamefont {K.}~\bibnamefont {Refson}},
  \bibinfo {author} {\bibfnamefont {M.}~\bibnamefont {Richter}}, \bibinfo
  {author} {\bibfnamefont {G.-M.}\ \bibnamefont {Rignanese}}, \bibinfo {author}
  {\bibfnamefont {S.}~\bibnamefont {Saha}}, \bibinfo {author} {\bibfnamefont
  {M.}~\bibnamefont {Scheffler}}, \bibinfo {author} {\bibfnamefont
  {M.}~\bibnamefont {Schlipf}}, \bibinfo {author} {\bibfnamefont
  {K.}~\bibnamefont {Schwarz}}, \bibinfo {author} {\bibfnamefont
  {S.}~\bibnamefont {Sharma}}, \bibinfo {author} {\bibfnamefont
  {F.}~\bibnamefont {Tavazza}}, \bibinfo {author} {\bibfnamefont
  {P.}~\bibnamefont {Thunstrom}}, \bibinfo {author} {\bibfnamefont
  {A.}~\bibnamefont {Tkatchenko}}, \bibinfo {author} {\bibfnamefont
  {M.}~\bibnamefont {Torrent}}, \bibinfo {author} {\bibfnamefont
  {D.}~\bibnamefont {Vanderbilt}}, \bibinfo {author} {\bibfnamefont {M.~J.}\
  \bibnamefont {van Setten}}, \bibinfo {author} {\bibfnamefont {V.~V.}\
  \bibnamefont {Speybroeck}}, \bibinfo {author} {\bibfnamefont {J.~M.}\
  \bibnamefont {Wills}}, \bibinfo {author} {\bibfnamefont {J.~R.}\ \bibnamefont
  {Yates}}, \bibinfo {author} {\bibfnamefont {G.-X.}\ \bibnamefont {Zhang}}, \
  and\ \bibinfo {author} {\bibfnamefont {S.}~\bibnamefont {Cottenier}},\ }\href
  {\doibase 10.1126/science.aad3000} {\bibfield  {journal} {\bibinfo  {journal}
  {Science}\ }\textbf {\bibinfo {volume} {351}},\ \bibinfo {pages} {aad3000}
  (\bibinfo {year} {2016})}\BibitemShut {NoStop}%
\bibitem [{pbs()}]{pbspro}%
  \BibitemOpen
  \href@noop {} {}\bibinfo {howpublished}
  {http://www.pbsworks.com/Product.aspx?id=1}\BibitemShut {NoStop}%
\bibitem [{slu()}]{slurm}%
  \BibitemOpen
  \href@noop {} {}\bibinfo {howpublished}
  {https://computing.llnl.gov/linux/slurm/}\BibitemShut {NoStop}%
\bibitem [{sge()}]{sge}%
  \BibitemOpen
  \href@noop {} {}\bibinfo {howpublished}
  {https://www.oracle.com/technetwork/oem/grid-engine-166852.html}\BibitemShut
  {NoStop}%
\bibitem [{tor()}]{torque}%
  \BibitemOpen
  \href@noop {} {}\bibinfo {howpublished}
  {http://www.adaptivecomputing.com/products/open-source/torque/}\BibitemShut
  {NoStop}%
\bibitem [{aii()}]{aiida_plugin_registy}%
  \BibitemOpen
  \href@noop {} {}\bibinfo {howpublished}
  {https://aiidateam.github.io/aiida-registry/}\BibitemShut {NoStop}%
\bibitem [{pip()}]{pip}%
  \BibitemOpen
  \href@noop {} {}\bibinfo {howpublished}
  {https://pip.pypa.io/en/stable/}\BibitemShut {NoStop}%
\bibitem [{\citenamefont {Talirz}\ \emph {et~al.}(2020)\citenamefont {Talirz},
  \citenamefont {Kumbhar}, \citenamefont {Passaro}, \citenamefont {Yakutovich},
  \citenamefont {Granata}, \citenamefont {Gargiulo}, \citenamefont {Borelli},
  \citenamefont {Uhrin}, \citenamefont {Huber}, \citenamefont {Zoupanos},
  \citenamefont {Adorf}, \citenamefont {Andersen}, \citenamefont
  {Sch{\"{u}}tt}, \citenamefont {Pignedoli}, \citenamefont {Passerone},
  \citenamefont {VandeVondele}, \citenamefont {Schulthess}, \citenamefont
  {Smit}, \citenamefont {Pizzi},\ and\ \citenamefont {Marzari}}]{Talirz2020}%
  \BibitemOpen
  \bibfield  {author} {\bibinfo {author} {\bibfnamefont {L.}~\bibnamefont
  {Talirz}}, \bibinfo {author} {\bibfnamefont {S.}~\bibnamefont {Kumbhar}},
  \bibinfo {author} {\bibfnamefont {E.}~\bibnamefont {Passaro}}, \bibinfo
  {author} {\bibfnamefont {A.~V.}\ \bibnamefont {Yakutovich}}, \bibinfo
  {author} {\bibfnamefont {V.}~\bibnamefont {Granata}}, \bibinfo {author}
  {\bibfnamefont {F.}~\bibnamefont {Gargiulo}}, \bibinfo {author}
  {\bibfnamefont {M.}~\bibnamefont {Borelli}}, \bibinfo {author} {\bibfnamefont
  {M.}~\bibnamefont {Uhrin}}, \bibinfo {author} {\bibfnamefont {S.~P.}\
  \bibnamefont {Huber}}, \bibinfo {author} {\bibfnamefont {S.}~\bibnamefont
  {Zoupanos}}, \bibinfo {author} {\bibfnamefont {C.~S.}\ \bibnamefont {Adorf}},
  \bibinfo {author} {\bibfnamefont {C.~W.}\ \bibnamefont {Andersen}}, \bibinfo
  {author} {\bibfnamefont {O.}~\bibnamefont {Sch{\"{u}}tt}}, \bibinfo {author}
  {\bibfnamefont {C.~A.}\ \bibnamefont {Pignedoli}}, \bibinfo {author}
  {\bibfnamefont {D.}~\bibnamefont {Passerone}}, \bibinfo {author}
  {\bibfnamefont {J.}~\bibnamefont {VandeVondele}}, \bibinfo {author}
  {\bibfnamefont {T.~C.}\ \bibnamefont {Schulthess}}, \bibinfo {author}
  {\bibfnamefont {B.}~\bibnamefont {Smit}}, \bibinfo {author} {\bibfnamefont
  {G.}~\bibnamefont {Pizzi}}, \ and\ \bibinfo {author} {\bibfnamefont
  {N.}~\bibnamefont {Marzari}},\ }\href {http://arxiv.org/abs/2003.12510} {\ ,\
  \bibinfo {pages} {1} (\bibinfo {year} {2020})},\ \Eprint
  {http://arxiv.org/abs/2003.12510} {arXiv:2003.12510} \BibitemShut {NoStop}%
\bibitem [{Note2()}]{Note2}%
  \BibitemOpen
  \bibinfo {note} {\protect \url {https://www.postgresql.org/}}\BibitemShut
  {NoStop}%
\bibitem [{Note3()}]{Note3}%
  \BibitemOpen
  \bibinfo {note} {\protect \url {https://www.rabbitmq.com/}}\BibitemShut
  {NoStop}%
\bibitem [{cir()}]{circus}%
  \BibitemOpen
  \href@noop {} {}\bibinfo {howpublished}
  {https://circus.readthedocs.io/}\BibitemShut {NoStop}%
\bibitem [{\citenamefont {Uhrin}\ and\ \citenamefont
  {Huber}(2020)}]{Uhrin2020}%
  \BibitemOpen
  \bibfield  {author} {\bibinfo {author} {\bibfnamefont {M.}~\bibnamefont
  {Uhrin}}\ and\ \bibinfo {author} {\bibfnamefont {S.~P.}\ \bibnamefont
  {Huber}},\ }\href {http://arxiv.org/abs/2005.07475} {\bibfield  {journal}
  {\bibinfo  {journal} {arXiv:2005.07475 [cs]}\ } (\bibinfo {year} {2020})},\
  \bibinfo {note} {arXiv: 2005.07475}\BibitemShut {NoStop}%
\bibitem [{Note4()}]{Note4}%
  \BibitemOpen
  \bibinfo {note} {\protect \url
  {https://aio-pika.readthedocs.io/en/latest/}}\BibitemShut {NoStop}%
\bibitem [{Note5()}]{Note5}%
  \BibitemOpen
  \bibinfo {note} {\protect \url
  {https://aiidateam.github.io/aiida-registry/}}\BibitemShut {NoStop}%
\end{thebibliography}%
